\documentclass[amsmath,amssymb,twocolumn,notitlepage,footinbib,superscriptaddress,floatfix,eprint,aps,physrev,10pt,tightenlines]{revtex4-2}

\PassOptionsToPackage{pdftex,
pdfencoding=auto,
pdfnewwindow=true,
pdfusetitle=true,
bookmarks=true,
bookmarksnumbered=true,
bookmarksopen=true,
pdfpagemode=UseThumbs,
bookmarksopenlevel=1,
pdfpagelabels=true,
breaklinks=true,
colorlinks=true,
}{hyperref}
\usepackage{hyperref}
\usepackage{amsmath, amssymb, amsthm, physics}
\usepackage[utf8]{inputenc}
\usepackage[countmax]{subfloat}
\usepackage{comment}
\usepackage{todonotes}
\usepackage{bbm}
\usepackage{graphicx}
\usepackage{framed}
\usepackage{color}
\usepackage{soul}
\usepackage{epstopdf}
\usepackage{dsfont}
\usepackage{wrapfig}
\usepackage[caption=false]{subfig}
\usepackage{relsize}
\usepackage{subcaption}
\newcommand{\powerset}{\raisebox{.15\baselineskip}{\Large\ensuremath{\wp}}}

\definecolor{npronk}{rgb}{0.3,0.5,0}

\newtheorem{theorem}{Theorem}

\begin{document}
\title{
Quantum violation of Causal Friendliness in the instrumental scenario}


\author{Emanuele Polino}
\email{e.polino@griffith.edu.au , emanuele.polino@gmail.com}
\affiliation{Queensland Quantum and Advanced Technologies Research Institute, Griffith University, Yuggera Country, Brisbane, QLD 4111, Australia}

\author{Giulia Lelli} 
\affiliation{Dipartimento di Fisica - Sapienza Universit\`{a} di Roma, P.le Aldo Moro 5, I-00185 Roma, Italy}

\author{Aaditya Datar} 
\affiliation{Queensland Quantum and Advanced Technologies Research Institute, Griffith University, Yuggera Country, Brisbane, QLD 4111, Australia}

\author{Marwan Haddara}
\affiliation{Queensland Quantum and Advanced Technologies Research Institute, Griffith University, Yugambeh Country, Gold Coast, QLD 4222, Australia}

\author{Davide Poderini}

\affiliation{Università degli Studi di Pavia, Dipartimento di Fisica, QUIT Group, via Bassi 6, 27100 Pavia, Italy}

\author{Giorgio Milani} 

\affiliation{Dipartimento di Fisica - Sapienza Universit\`{a} di Roma, P.le Aldo Moro 5, I-00185 Roma, Italy}

\author{Sergei Slussarenko} 

\affiliation{Queensland Quantum and Advanced Technologies Research Institute, Griffith University, Yuggera Country, Brisbane, QLD 4111, Australia}

\author{Howard M. Wiseman} 

\affiliation{Queensland Quantum and Advanced Technologies Research Institute, Griffith University, Yuggera Country, Brisbane, QLD 4111, Australia}

\author{Nora Tischler} 
\affiliation{Queensland Quantum and Advanced Technologies Research Institute, Griffith University, Yuggera Country, Brisbane, QLD 4111, Australia}

\author{Eric G. Cavalcanti} 
\email{e.cavalcanti@griffith.edu.au}
\affiliation{Queensland Quantum and Advanced Technologies Research Institute, Griffith University, Yugambeh Country, Gold Coast, QLD 4222, Australia}

\author{Fabio Sciarrino}
\email{fabio.sciarrino@uniroma1.it}

\affiliation{Dipartimento di Fisica - Sapienza Universit\`{a} di Roma, P.le Aldo Moro 5, I-00185 Roma, Italy}

\begin{abstract}
How far can quantum theory depart from a classical worldview? Bell inequality violations show that no theory can simultaneously uphold absoluteness of observed events, freedom of choice, relativistic causality, and classical causal explanations. While generalized notions of causality, capable of incorporating quantum resources, can account for Bell correlations, the Local Friendliness theorem showed that, in extended Wigner’s friend scenarios, the conflict between quantum theory and causal explanations persists even after classical causality is abandoned. This revealed a new form of nonclassicality stronger than what was revealed by Bell’s theorem.
However, Local Friendliness has been confined to Bell causal structures so far, and it remains unknown whether a related form of nonclassicality persists in more general causal networks.
Here we answer this question by considering a Wigner's friend version of the instrumental network, a paradigmatic causal structure in which communication between parties is allowed, and, unlike the Bell scenario, relativistic causality does not impose any restriction.
We show, both theoretically and in a proof-of-principle photonic experiment, that quantum correlations are incompatible with Causal Friendliness, which is the conjunction of the absoluteness of observed events and the generalized nonclassical causal constraints imposed by the network. Such constraints extend relativistic causality from Bell scenarios to networks of arbitrary structure.
These results establish a form of quantum nonclassicality stronger than previously observed in instrumental-network violations of classical causality, generalizing Local Friendliness beyond Bell-Wigner scenarios and uncovering a new class of correlations with potential applications in future quantum technologies.
\end{abstract}

\maketitle

\section{Introduction}

A central goal of science is not only to predict observations but also to explain them in terms of underlying causal mechanisms. From physics to medicine, causal explanations provide the conceptual framework through which observed correlations are understood~\cite{pearl2009causality}. Quantum theory, however, has repeatedly challenged this programme. Bell's theorem~\cite{bell1964einstein,brunner2014bell} showed that no classical causal explanation can reproduce the observed correlations between distant quantum systems in the Bell causal structure, revealing a profound incompatibility between quantum theory and the classical causal worldview~\cite{wood2015lesson, wiseman2017causarum}. 
This incompatibility, conventionally described as quantum nonlocality~\cite{brunner2014bell}, invokes  the classical notion of causes. 
 In response to Bell's theorem, generalized causal frameworks have been proposed, allowing causal influences and common causes to be nonclassical, thereby accounting for correlations that elude classical causal models without abandoning the causal structure itself~\cite{wood2015lesson,chaves2015unifying,cavalcanti2014modifications,henson2014theory,costa2016quantum,allen2017quantum,wolfe2021quantum,barrett2019quantum}. 

More recently, however, the Local Friendliness (LF) theorem~\cite{bong2020strong} has shown that abandoning classical causality does not resolve the tension between quantum theory and causal explanation~\cite{cavalcanti2021implications,ying2024relating}. By combining the Bell scenario with a ``Wigner's friend'' and an assumption of Absoluteness of Observed Events (AOE), 
it demonstrates that quantum predictions are incompatible even with generalized causal explanations~\cite{bong2020strong,cavalcanti2021implications,ying2024relating}.

These developments leave open a fundamental question. Previous theoretical and experimental demonstrations~\cite{bong2020strong,haddara2025local,wiseman2023thoughtful,schmid2023review,ying2024relating} of this form of nonclassicality based on AOE and generalized causal constraints have been formulated within Bell causal structures, where distant observers perform measurements on systems distributed by a common source. Indeed, the original LF theorem used an assumption based on the relativistic causal arrow principle~\cite{cavalcanti2021implications} that does not impose constraints on scenarios without spacelike separation. 
While those constraints can be motivated from generalized causality in a Bell causal structure~\cite{ying2024relating}, it remains unknown whether they reflect a broader limitation of nonclassical causal explanations in general causal networks. 

\begin{figure*}[ht!]
  \includegraphics[width = 0.9\textwidth]{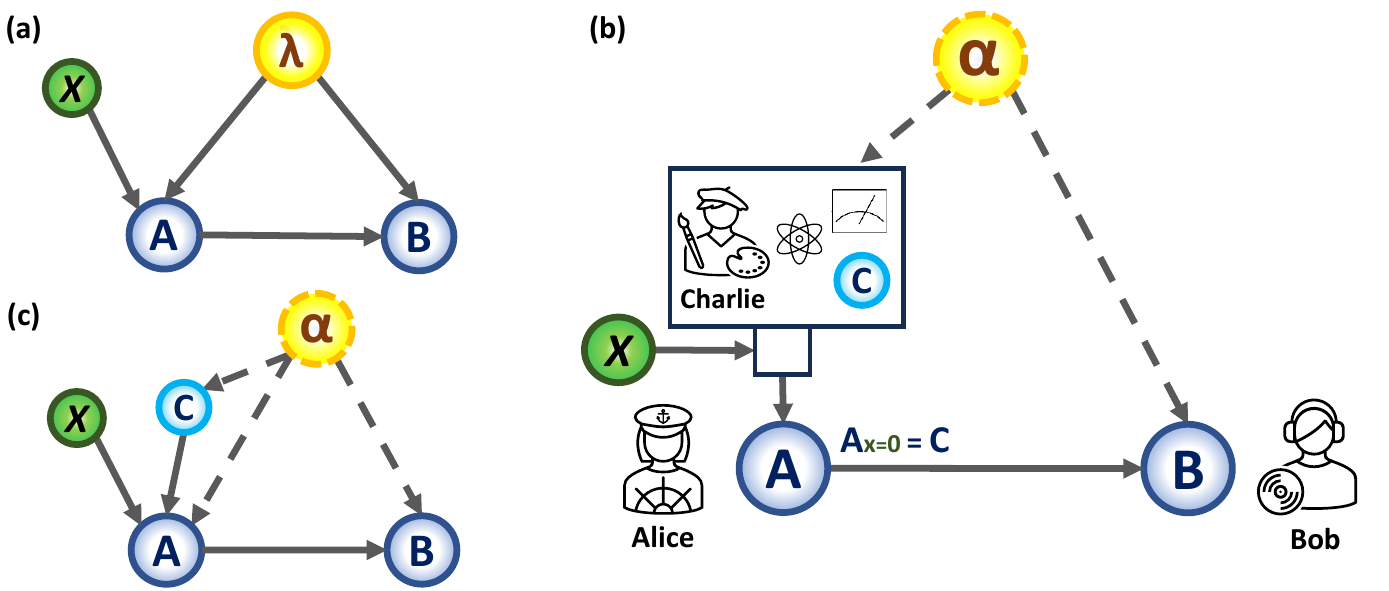}
\caption{
\textbf{(a)} Standard classical causal structure of the instrumental scenario, where a single (classical) source $\lambda$ shares correlations between the two measurement nodes $A$ and $B$, representing the observed variables of the network. The variable $A$ (Alice's measurement), influenced by the free choice $X$, can affect Bob's measurement variable $B$. 
\textbf{(b)} Extended Wigner's friend scenario built within the instrumental framework, where an outcome $C$ is produced after the interaction between the incoming subsystem and an observer system in an isolated laboratory. The laboratory is then measured by a superobserver with a measurement choice $X$ and outcome $A$. The assumptions in Eqs.~\eqref{eq:aoe}-\eqref{eq:aoe2} imply that, whenever $X=0$, $A=C$.
\textbf{(c)} Causal diagram of the variables involved in the Instrumental Wigner's friend scenario, where the presence of two channels from the (nonclassical) source $\alpha$ to the node $A$ is made explicit: one classical channel (acting when $X=0$) and one nonclassical channel (acting when $X \ne 0$). The dashed (solid) arrows represent the causal influences from a nonclassical (classical) node.}
\label{fig:instrumental}
\end{figure*}

Here, we show that indeed this form of nonclassicality is not confined to Bell causal structures. We do this by considering the instrumental scenario, a paradigmatic causal network 
with a central role in both causal inference and quantum networks~\cite{balke1997bounds,pearl1995testability,pearl2009causality,chaves2018quantum,van2019quantum,agresti2022experimental}. In the instrumental network, the outcome of one observer, Alice, can directly influence the measurement performed by another observer, Bob (see Fig.~\ref{fig:instrumental}), so spacelike separation plays no role and the relativistic causal arrow assumption of the Local Friendliness theorem no longer provides the relevant constraint. 
We introduce a Wigner's-friend extension of the instrumental scenario and formulate a set of theory-independent assumptions that we call \emph{Causal Friendliness} (CF). These assumptions combine AOE with constraints derived for the instrumental causal structure using generalized causal models~\cite{henson2014theory} based on Operational Probabilistic Theories (OPTs). This framework extends the description of causal relations beyond classical and quantum theory while remaining compatible with the no-signaling principle~\cite{chiribella2010probabilistic, henson2014theory}.

 A violation of CF, formalised in Sec.~\ref{sec:assumptions},  reveals a strong form of nonclassicality beyond the failure of classical causal models (CCMs), ruling out causal explanations even when the underlying resources and influences are allowed to be nonclassical. We prove that quantum theory violates CF in the instrumental scenario (Sec.~\ref{Section:StrongQuantumNonclassicalityintheInstrumentalScenario}) and demonstrate this violation in a proof-of-principle photonic experiment using polarization-entangled photons, active feed-forward, and a single-photon path to encode the friend's outcome (Sec.~\ref{sec:experiment}). The observed correlations violate a device-independent Bell-like CF inequality by more than $120$ standard deviations, without relying on a characterization of the internal operation of the experimental devices. 
These results reveal the tension between quantum theory and models that respect generalized causality in the assumed causal structure and the AOE. By taking this strong form of quantum nonclassicality beyond the Bell scenario and into a broader landscape of causal structures, our work opens new avenues for quantum foundations and quantum technologies.

\section{Causal Friendliness beyond the Bell scenario}

\subsection{Friendliness: nonclassicality beyond the violation of classical causality} 
Bell's theorem is traditionally understood as ruling out local hidden-variable theories~\cite{bell1964einstein,brunner2014bell}. 
From the modern perspective of causal inference~\cite{wood2015lesson,chaves2015unifying} however, it is viewed as the incompatibility between quantum correlations and four basic assumptions, which, following Refs.~\cite{wiseman2017causarum,cavalcanti2021implications}, are: (i) Absoluteness of Observed Events (AOE), (ii) relativistic causal arrow, (iii) freedom of choice (or independent interventions~\cite{cavalcanti2021implications}), and (iv) classical causal explanation. 
In particular, classical causal explanations are a much debated concept in physics~\cite{price2007causation,wood2015lesson,wiseman2017causarum,rovelli2025back}.   In the Bell scenario, the classical causal notion is embodied in \emph{Reichenbach's principle of common cause}~\cite{cavalcanti2014modifications, cavalcanti2021implications}: 
that for any two sets of correlated events {\bf A} and {\bf B}, such that no event in either is a cause of any event in the other, there exists a set of common causes {\bf C}, such that conditioning on {\bf C} eliminates the correlation. 
Reichenbach's principle follows from the more general \emph{classical causal Markov condition}~\cite{pearl2009causality}: that a set of events is independent of any non-effects given a sufficient specification of its direct causes. 
Violations of the classical causal Markov condition have been studied~\cite{tavakoli2022bell,branciard2012bilocal,fritz2012BellTheoremCorrelation,renou2019genuine,chaves2021causal,van2019quantum} and experimentally demonstrated~\cite{carvacho2017experimental,saunders2017experimental,chaves2018quantum,poderini2020experimental,wang2023certification,polino2023experimental,gu2023experimental,wang2026experimental} in various causal networks beyond the Bell scenario. These observations have motivated a broad research programme aimed at extending classical causal explanations to the quantum, or more generally nonclassical (OPT) regime,
to explain quantum nonclassicality~\cite{wood2015lesson,chaves2015unifying,henson2014theory,costa2016quantum,allen2017quantum,wolfe2021quantum,barrett2019quantum}.

However, existing frameworks of generalized causal explanations are challenged by the quantum measurement problem---the tension between the deterministic unitary evolution of quantum theory and the emergence of definite measurement outcomes~\cite{fine1970insolubility,bell1990against,wigner1984review,maudlin1995three,bub2010two,cavalcanti2016bell,brukner2017quantum,gisin2018collapse,ormrod2026theories} best illustrated by Wigner's friend scenarios~\cite{wigner1995remarks,wigner1997problem,deutsch1985quantum,frauchiger2018quantum}. When extended to quantum systems distributed among distant parties~\cite{brukner2017quantum,brukner2018no,proietti2019experimental}, these scenarios reveal that even generalized causal explanations, which correspond to relativistic causality in Bell scenarios,
become incompatible with quantum theory if AOE and free choices are retained~\cite{bong2020strong,wiseman2023thoughtful,cavalcanti2021implications,brukner2022wigner,schmid2023review,ying2024relating}. 
We stress that, although the LF assumptions were originally introduced in the context of extended Wigner's friend experiments, which provide a natural motivation for AOE, such scenarios represent only one possible perspective in which the CF assumptions may be considered. More generally, CF can be viewed as a theory-independent relaxation of classical causal constraints defining partially deterministic models, which also have application to quantum information protocols such as device-independent randomness certification~\cite{woodhead2014imperfections}. 

We now extend this framework beyond Bell scenarios, to the instrumental scenario, which therefore requires replacing Bell-specific relativistic causal arrow constraints with generalized causal constraints. 

\subsection{Causal Friendliness in the Instrumental Scenario}
\label{sec:assumptions}
The (classical) instrumental causal structure, shown in Fig.~\ref{fig:instrumental}(a), consists of a latent classical common cause $\Lambda$ acting on $A$ and $B$, a direct influence from $A$'s outcome to $B$, and an ``instrumental'' variable (here, the measurement setting) $X$ that controls $A$. It can be viewed as a relaxed Bell scenario in which one-way communication of outcomes from $A$ to $B$ is allowed, and only $A$ is influenced by a free setting. 
Instrumental tests are central to causal inference because they enable us to distinguish between the two possible classical explanations for the correlations between $A$ and $B$: a common cause  or a direct cause~\cite{balke1997bounds,pearl1995testability,pearl2009causality}. For this reason, this scenario is widely applied across scientific disciplines~\cite{angrist1996identification,mogstad2018using}.

In the classical causal framework, the Markov condition~\cite{pearl2009causality} applied to the instrumental network imposes the following constraint on the probabilities: $P(a,b|x)=\sum_\lambda P(a|x,\lambda)\, P(b|a,\lambda)\,P(\lambda)$. 
This constraint leads to inequalities~\cite{bonet2001instrumentality} that can be violated by quantum systems, as shown theoretically and experimentally in Ref.~\cite{chaves2018quantum}. A connection between violations of standard Bell and instrumental inequalities has also been established~\cite{van2019quantum}. The instrumental scenario has further been explored in the contexts of post-quantum steering~\cite{sainz2020bipartite,cavalcanti2022post}, as well as forms of nonclassicality enabled by interventions, i.e., actions assumed to be able to force the variable $A$ to assume certain values independently of its causal parents~\cite{gachechiladze2020quantifying,agresti2022experimental}.

Now consider Fig.~\ref{fig:instrumental}(b), illustrating a generalization of the instrumental scenario to one involving Wigner's friend. 
The subsystem in Alice’s station is first measured by her friend Charlie inside an isolated laboratory. Alice acts as a superobserver outside the laboratory and can perform quantum measurements on the joint system composed of Charlie and the subsystem. Charlie always performs a fixed measurement on the subsystem, yielding an outcome denoted by $C$. 
For one of Alice’s measurement settings, say $X=0$, she simply opens the laboratory and asks Charlie for the value of $C$, and records $A=C$. For all other values of $X$, Alice performs a different measurement on the composite Charlie--subsystem laboratory. Note that, under the assumption of AOE (Absoluteness of Observed Events, stated rigorously below), $C$ is always as real as $A$. However, this is not tantamount to assuming that Charlie's measurement collapses the quantum wavefunction; there are quantum interpretations where certain properties have well-defined values even when the quantum state is not in the associated eigenstate~\cite{bohm2006undivided},
and the no-go theorem has to apply to all conceivable theories satisfying the theorem's assumptions.

So far, we have described one station, which consists of one superobserver (Alice) and one friend (Charlie). There is another station, with a regular observer (Bob), but, unlike in Refs.~\cite{bong2020strong,cavalcanti2021implications,wiseman2023thoughtful}, these stations are not subject to the usual bipartite Bell causal structure. Rather, in this instrumental scenario,   after completing her measurement, Alice sends her outcome to Bob’s station, where the second subsystem is measured in a way that may depend on Alice’s result,
and yields an outcome $B$. Following the framework of~\cite{ying2024relating}, the generalized causal structure 
corresponding to this scenario is depicted in Fig.~\ref{fig:instrumental}(c): the  subsystems held by the two stations share a {\em nonclassical} common cause $\alpha$. The dashed arrows thus represent nonclassical channels while the solid ones are classical channels;  $X$, $C$, $A$, and $B$ are observed (classical) variables.

Given this scenario, we now list and explain the CF assumptions that constrain the observed probability distributions $\powerset(a,b|x)$: 

  \textbf{(1)} \textbf{Absoluteness of Observed Events (AOE)}. AOE imposes that each observed event has a value that is absolute, i.e., not relative to anything (for discussions on this assumption in extended Wigner's friend experiments, see Refs.~\cite{cavalcanti2021implications,cavalcanti2021view,wiseman2023thoughtful,adlam2024does,frauchiger2018quantum,schmid2023review}). 
  In our scenario, this implies the existence of a theoretical joint probability distribution $P(a,c,b|x)$ that reproduces the observed probability $\powerset (a,b|x)$: 
\begin{equation}
\label{eq:aoe}
\powerset(a,b|x)=\sum_{c} P(a,c,b|x)\; \;\forall \; a,b,x\, .
\end{equation}

The description of the scenario for the setting $X=0$ also motivates the auxiliary \emph{tracking assumption}~\cite{schmid2023review}, that when Alice queries Charlie’s outcome, the value he reports corresponds to the value $C$ that is assumed to exist by AOE:
\begin{equation}
\label{eq:aoe2}
P(a|c,x=0)=\delta_{a,c}\; \;\forall\, a,c \;.
\end{equation}

An equivalent constraint to the perfect tracking in Eq.~\eqref{eq:aoe2} can nevertheless be obtained from a weaker assumption. One may indeed allow imperfect correlations between $A$ and $C$—for example, due to classical noise in the friend’s measurement or report—without introducing any additional nonclassical correlations between $A$ and $B$.
To cover these more general situations, the tracking assumption may be formulated more broadly by requiring only that, when $X=0$, the information communicated from Charlie to Alice is related to $A$ through a classical channel [solid arrow between $C$ and $A$ in Fig.~\ref{fig:instrumental}(c)], with no nonclassical side channel [dashed arrow between $\alpha$ and $A$ in Fig.~\ref{fig:instrumental}(c)]. We refer to this screening-off assumption as a \emph{reliable friend}.
Importantly, irrespective of the underlying description, whether based on the tracking assumption or on a classical-channel model of the reliable friend, Eq.~\eqref{eq:aoe2} provides the mathematical constraint relevant to CF~\footnote{Allowing a stochastic classical reporting channel does not allow correlations that are not reproducible by conditional determinism. Indeed, any classical channel $T(a|c)$ can be represented as a deterministic function $a=f(c,r)$ of its input and an auxiliary classical random variable $R$. Absorbing this randomness into an enlarged latent description therefore makes Alice's response at $x=0$ deterministic. Defining the effective classical record $C'=f(C,R)$ recovers the tracking relation $P(a|x=0,c')=\delta_{a,c'}$, although $C'$ need not coincide with Charlie's original record $C$.}. 
As shown in the following sections and discussed in Appendix~\ref{app:relaxationofAOE}, the reliable-friend assumption can be further relaxed in a distinct way by allowing a limited nonclassical contribution to the channel from the source to $A$ when $X=0$. This weakens the constraint in Eq.~\eqref{eq:aoe2} and, consequently, CF as a whole.

\textbf{(2)} \textbf{Generalized causality constraints in the instrumental scenario.} These constraints are derived from the OPT generalized Markov condition introduced by Ref.~\cite{henson2014theory}. This condition is expressed in an abstract diagrammatic framework and represents a relaxation of classical causality, allowing nonclassical causal relations between variables in a theory-independent way, and extending the no-signaling condition to arbitrary nonclassical causal structures. We refer to Appendix~\ref{app:OPTs} and Ref.~\cite{henson2014theory} for more details.

We first consider the constraint arising from generalized causality applied to the instrumental causal structure, which reflects the absence of a directed causal path from $X$ to $C$, and which is motivated by choosing $X$ via suitable external variables. Commonly called \textit{Freedom of Choice}, or \emph{No-Superdeterminism}, it requires statistical independence between Alice's choice $X$ and the variable $C$: 
    \begin{equation}
      \label{eq:frewill}
      P(x|c)=P(x) \;\; \;\forall \; x, c  \;\;.  
    \end{equation}

Next, we consider the constraints associated with the absence of a direct causal arrow from $X$ to $B$. 
In the instrumental scenario, the OPT causality constraints allow for the possibility that $X$ can signal to $B$, due to the existence of a causal path via $A$, so that $\powerset(b|x)\ne \powerset(b)$ in general. Nevertheless, the absence of a direct causal arrow from $X$ to $B$, together with the freedom-of-choice constraint above, imposes the following inequality constraints (details in Appendix~\ref{app:pearlsunique})~\cite{henson2014theory}:
         \begin{equation}
      \begin{split}
      \label{eq:locality}
  P(a,b=-1|x,&c) + P(a,b=+1|x',c)\le 1\;\\
 &\forall \; x\ne x',\;a\;, c\;.
   \end{split}
    \end{equation}

These inequalities are identical, except for the additional conditioning on $c$, to the standard \emph{Pearl inequalities}~\cite{pearl1995testability}, known to hold in both classical and OPT instrumental scenarios~\cite{henson2014theory} (see Appendix~\ref{app:pearlsunique}).
We note that, for the particular case of the instrumental scenario, these conditions are mathematically equivalent to the constraints arising from the exclusivity principle~\cite{cabello2012specker,cabello2013simple} used in the graph-theoretic approach to noncontextual, local hidden variable models, and also in networks with hybrid assumptions~\cite{rodari2023characterizing} (see Appendix~\ref{app:ExclusivityrelationOPT}).
Together, Eqs.~\eqref{eq:aoe}--\eqref{eq:locality} define Causal Friendliness~\footnote{Note that the notion of Causal Friendliness introduced here is distinct from the ``Causal Time-Symmetric Friendliness'' considered in Ref.~\cite{mukherjee2026limits}, which combines AOE with other assumptions named Axiological Time Symmetry, No Retrocausality, and Screening via Pseudo Events in a sequential prepare-and-measure configuration. }.

\section{Quantum violation of Causal Friendliness in the instrumental scenario\label{Section:StrongQuantumNonclassicalityintheInstrumentalScenario}}

The assumptions in Eqs.~(\ref{eq:aoe})--(\ref{eq:locality}) defining CF give rise to partially deterministic behaviours~\cite{bong2020strong,haddara2025properties}, whereby deterministic outcomes occur only for a subset of inputs (here, $X=0$). For all the remaining inputs, the outcome statistics are constrained solely by the generalized causality principle. 
We consider the minimal scenario where  a quantum-classical gap can be found~\cite{chaves2018quantum,van2019quantum}: dichotomic outputs $A,B\in\{-1,+1\}$, and three choices of measurements $X\in \{0,1,2\}$.  Equations~(\ref{eq:aoe})-(\ref{eq:aoe2}) imply 
the output $A$ is deterministic when the measurement choice is $X=0$.
We fully characterize the resulting CF correlations by finding the polytope resulting from the constraints \eqref{eq:aoe}--\eqref{eq:locality}. 
In general, the following strict inclusion relations between various sets of correlations hold: $CCM \subset CF \subset OPT$, and $CCM \subset Q \subset OPT$. Here, the sets correspond to classical causal models $ CCM$,  Causal Friendliness $CF$, quantum $Q$, and operational probabilistic theories $OPT$.   
Note that the relation between the quantum set $Q$ and the Causal Friendliness set $CF$ is not one of inclusion; the two are distinct, intersecting sets.
The $CF$ instrumental polytope has 36 vertices and 48 facets. Among these facets, 8 are genuine CF, in the sense that they are not facets of the CCM nor of the OPT polytopes.

The 8 genuine CF facets, up to relabeling of the outcome values of the variables $A,\, B$ and the two non-deterministic input values of $X$ in the observed probability $\powerset(a,b|x)$, are described by a single class of genuine CF inequalities: 
\begin{equation}
     I_{\text{inst}} \equiv \langle AB \rangle_1-\langle AB \rangle_0+ \langle B \rangle_0 - \langle B \rangle_2  \overset{CF}{\le} 2 \;.
     \label{eq:genuinelf}
\end{equation}
The recorded outcomes are binary outcomes taking values $a,b\in\{-1,+1\}$, and the sufficient statistics are the mean of Bob’s outcome, 
$\langle B\rangle_x
=\sum_{a,b} b\, \cdot\powerset(a,b|x)$,
and the mean product of the outcomes, 
$\langle AB\rangle_x =\sum_{a,b}a\,\cdot b\,\cdot\powerset(a,b|x)$.

\begin{figure*}[ht!] \centering \subfloat[]{ \includegraphics[width=0.706\textwidth]{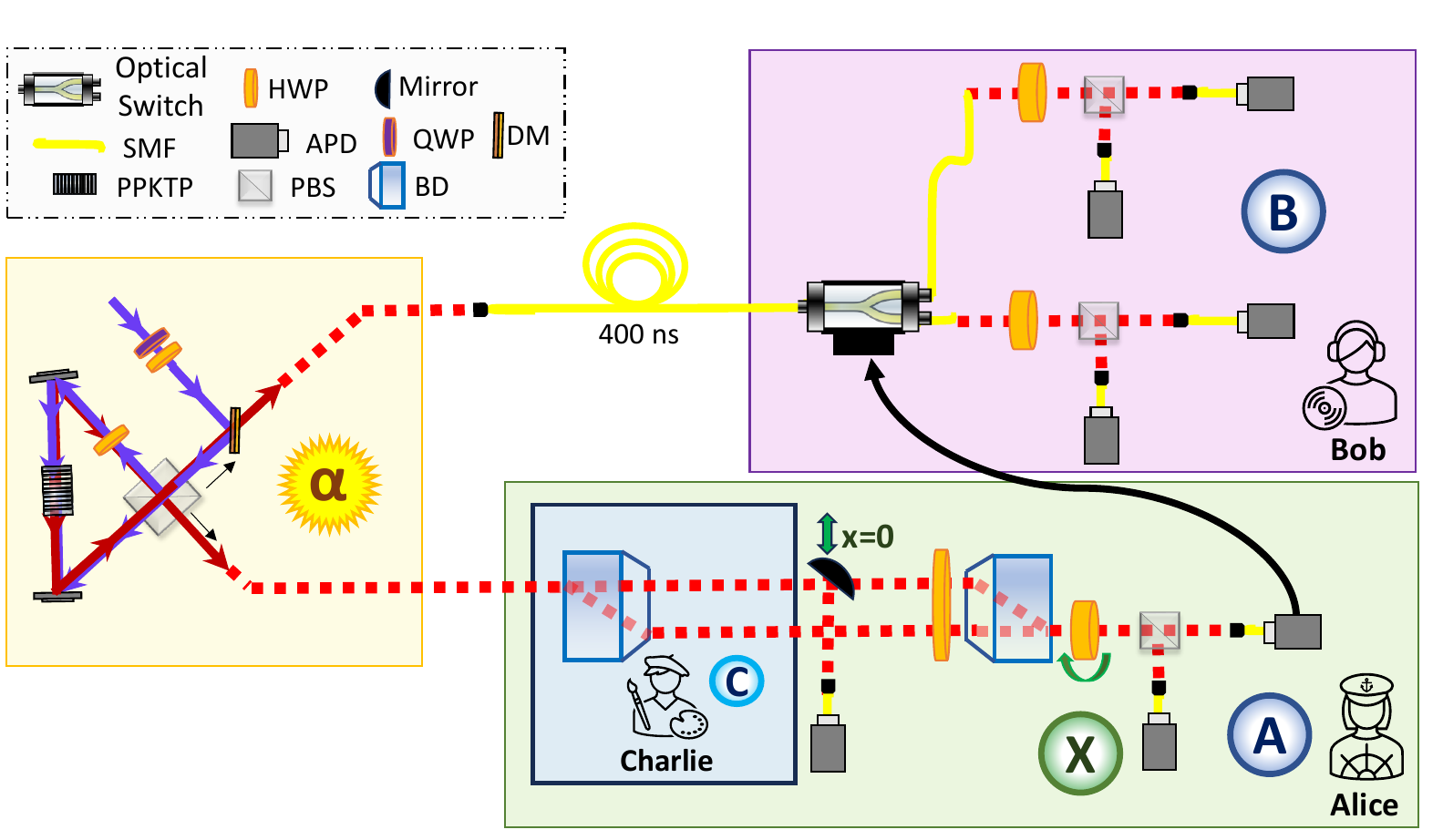} \label{fig:setup} }
\subfloat[]{ \includegraphics[width=0.273\textwidth]{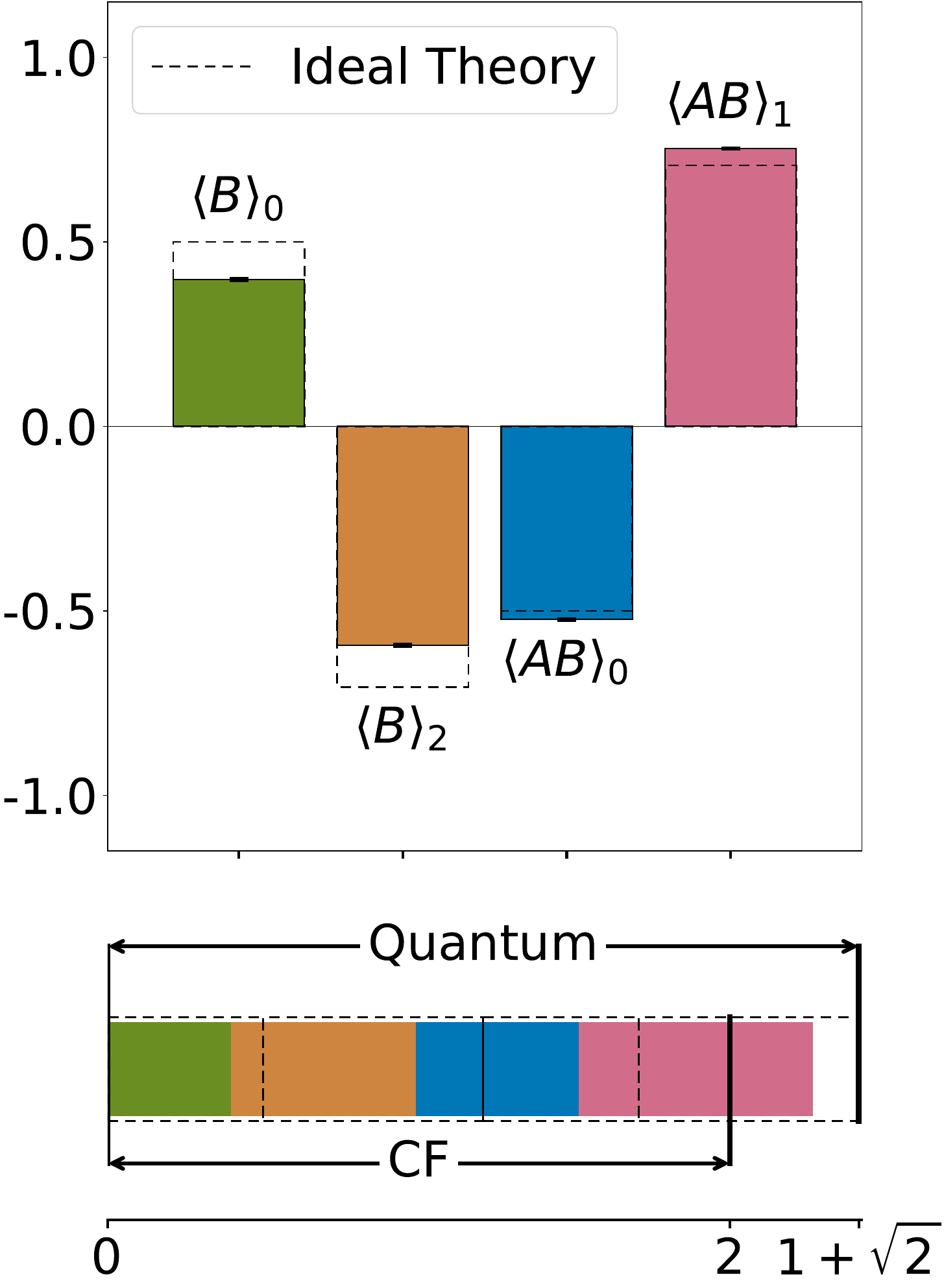} \label{fig:violationplot} } \caption{\textbf{Experimental setup and results.} \textbf{(a)} A polarization-entangled photon source generates the singlet state via a type-II SPDC process in a Sagnac interferometer. One photon is sent to Alice's station, where it undergoes the friend interaction and the final superobserver's measurements, whose settings are labeled by $X$. The other photon goes through a delay line and is then directed to one of Bob's measurement stations by an optical switch using a feed-forward signal based on Alice's measurement outcome.  APD = avalanche photodiode single-photon detector; BD = beam displacer; DM = dichroic mirror; HWP = half-wave plate; PBS = polarizing beam splitter; PPKTP = periodically poled potassium titanyl phosphate; QWP = quarter-wave plate; SMF = single-mode fiber. \textbf{(b)} Experimental correlator values estimated from the experimental statistics that violate the genuine CF facet in Eq.~\eqref{eq:genuinelf}. The error bars (barely visible) represent $3\sigma$, where the standard error $\sigma$ on the mean values is calculated by repeating the measurement $100$ times. The dashed grey lines indicate the values of the correlators calculated from an ideal maximally entangled singlet state and optimal measurements. The linear combination of the correlators generates the violation of the CF bound as depicted in the bottom part of the figure.} \label{fig:expsetup} \end{figure*}

Remarkably, quantum systems using two-qubit entangled states and projective measurements can violate Eq.~\eqref{eq:genuinelf}, showing that Causal Friendliness can be disproven in the instrumental scenario. 
To see this, consider the maximally entangled state: $\ket{\Psi_{AB}} = (\ket{01} -\ket{10})/\sqrt{2}$. 
Alice performs a binary projective measurement specified by an observable $\hat A_x$, while Bob measures an observable $\hat B_y$, with outcomes $a,b\in\{-1,+1\}$. Bob's setting is determined by Alice’s outcome, and conventionally we set $y=a$. The corresponding outcome projectors are $\Pi^A_{a|x}
=(\mathbb I_A+a\, \cdot \hat A_x)/2$, $\Pi^B_{b|y}
=(\mathbb I_B+b\, \cdot \hat B_y)/2$.  
The adaptive protocol therefore has joint measurement effects:
\begin{equation}
E_{ab|x}
=\Pi^A_{a|x}\otimes\Pi^B_{b|a},
\end{equation}
  yielding probabilities
$\powerset(a,b|x)
=\langle\Psi_{AB}|E_{ab|x}|\Psi_{AB}\rangle$. 
For each $x$, these effects satisfy 
$\sum_{a,b}E_{ab|x}=\mathbb I_{AB}$.
Choosing observables  
$$\hat A_0=-\sigma_z,\;
\hat A_1=(\sigma_z-\sigma_x)/\sqrt2,\;\;
\hat A_2=(\sigma_z+\sigma_x)/\sqrt2$$ for Alice, and $$\hat B_{+1}=\sigma_x,\;
\hat B_{-1}=-\sigma_z$$ for Bob, 
yields  $I_{\mathrm{inst}}=1+\sqrt2>2$, thereby violating the CF bound. 
We also prove that this is an optimal choice by showing that the achieved value for the violation matches the numerical upper bound obtained at the second level of the Navascués–Pironio–Acín hierarchy \cite{navascues2008convergent}.
The OPT bound for Eq.~\eqref{eq:genuinelf} is instead 3. 

These findings establish a central result of this work:
\begin{theorem}
Quantum predictions in the instrumental scenario are incompatible with constraints implied by Causal Friendliness.
\end{theorem}

In the Wigner's friend narrative, to observe a quantum violation of Causal Friendliness, the superobserver must coherently control both the friend and the relevant environment, while the friend’s role as a genuine observer must remain well justified~\cite{wiseman2023thoughtful}. We note, however, that the Wigner's friend narrative represents only a framework for CF, and the theorem and the inequalities found are fully theory-independent and ``narrative''-independent, since they follow from the mathematical assumptions in Eqs.~(\ref{eq:aoe})-(\ref{eq:locality}). These represent a strict relaxation of the standard CCM assumptions, and thus CF violation implies a notion of nonclassicality stronger than the violation of classical causal models. 


 
It is interesting to note that we have quantum violations of genuine CF inequalities with just three measurement settings, i.e., $|X|=3$, when $A$ and $B$ are dichotomic. 
That is, no additional measurement settings are required to violate the assumptions of CF beyond those needed to violate standard classical causal constraints. Note that we are not considering scenarios where one can use interventions that would allow one to violate classical causal constraints with only two measurements~\cite{gachechiladze2020quantifying,agresti2022experimental}. 
This differs from the standard Bell scenario, where demonstrating a genuine violation of LF necessitates a larger set of measurement inputs than those sufficient to violate local causality~\cite{bong2020strong}.

\section{Experimental demonstration}\label{sec:experiment}
We perform a photonic experiment implementing the Wigner's friend instrumental scenario to test, through quantum correlations, the constraints imposed by CF.
We generate polarization-entangled photon pairs from a spontaneous parametric down-conversion (SPDC) source based on a Sagnac interferometer geometry~\cite{fedrizzi2007wavelength}. One of the photons is sent to Alice's station, where   a   coupling between the path and the polarization is realized through a beam displacer (see Fig.~\ref{fig:expsetup}\subref{fig:setup}). This coupling, in the Wigner's friend narrative, represents the 
(reversible) measurement of the system, encoded in the photon's polarization, by the friend, encoded in the photon's path.
When the choice of the superobserver is $X=0$, a motorized mirror is inserted, and the path of the photon is directly measured by single-photon detectors in the two paths coupled with the horizontal or vertical polarization, respectively. If the measurement choice is $X \ne 0$, the path--polarization coupling is reversed by a second beam displacer, followed by the implementation of a polarization measurement.
The second photon is first delayed by about $400$ ns and then sent to Bob's station. 
There, the photon enters a fast optical switch that routes it to one of two measurement stations conditioned on Alice’s outcome: a click in the $+1$ detector triggers an electronic feed-forward signal that changes the selected output port of Bob's photon, whereas a click in the $-1$ detector leaves the switch in its default configuration.
After each of the two output ports of the switch, a static polarization measurement is performed followed by single-photon detection. Through this feed-forward, the output bit from Alice's measurement determines Bob's measurement. (See Appendix~\ref{app:expdetails} for experimental details and the outcome assignment strategy.)

\begin{figure}[ht!]
  \centering
  \includegraphics[width = 1.01\columnwidth]{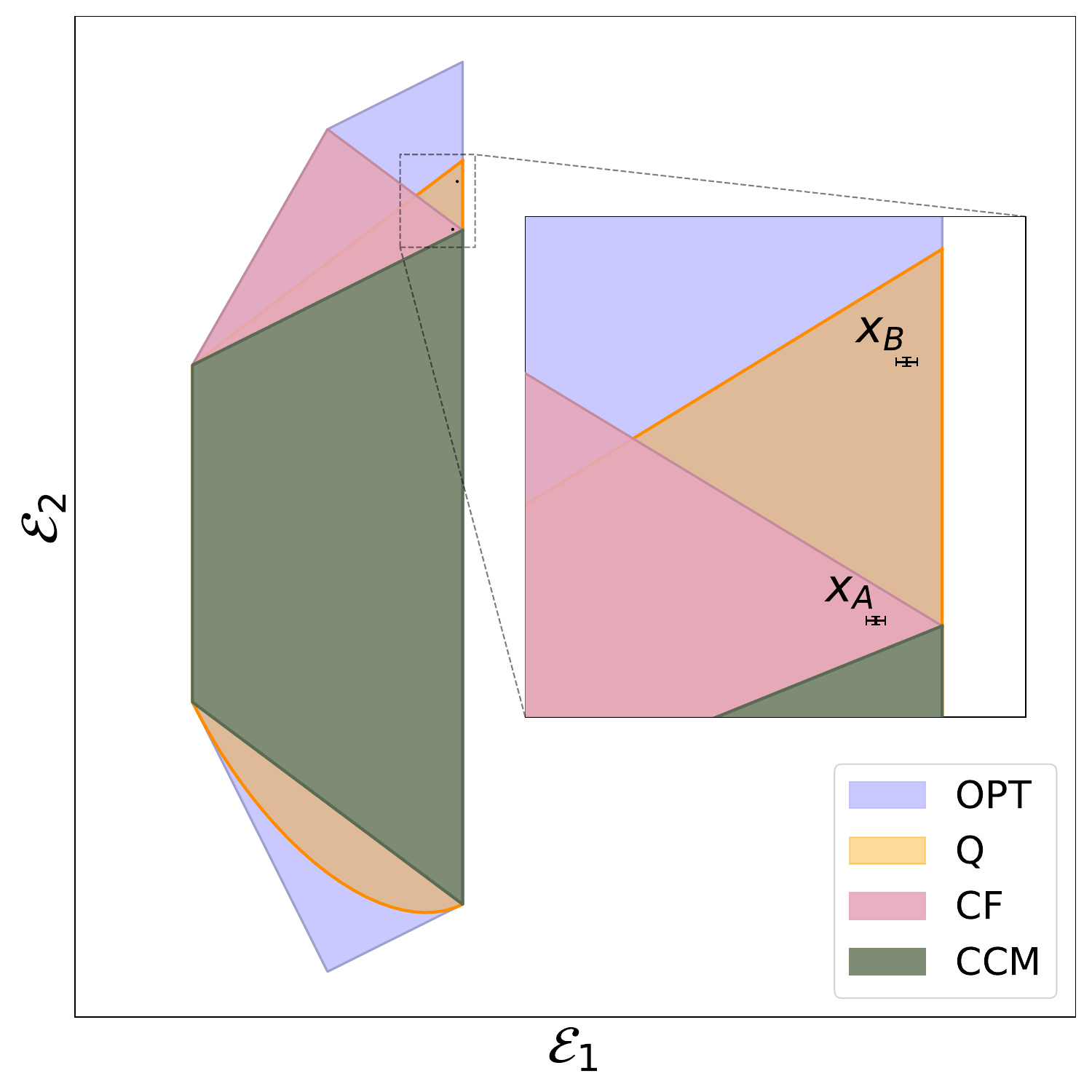}
\caption{\textbf{Correlation sets.} 2-D projection of correlation sets corresponding to the Operational Probabilistic Theories (OPT), quantum theory (Q), Causal Friendliness (CF), and Classical Causal Models (CCM). The experimental data are plotted as points to visualize the violations they imply: the point $X_B$ lies outside both the CF and CCM polytopes, whereas the point $X_A$, obtained from noisier measurements, lies entirely within the CF polytope but outside the CCM polytope. The error bars represent $3\sigma$, where $\sigma$ is the standard error of the mean value in $100$ repetitions. The expressions $\mathcal{E}_1$ and $\mathcal{E}_2$ on the axes are linear combinations of specific correlations; further details are provided in Appendix~\ref{app:plotexpressions}.}
\label{fig:results}
\end{figure}

We begin by testing the genuine CF facet in Eq.~\eqref{eq:genuinelf}, and we measure $I_{\text{inst}}^{\mathrm{exp}} = 2.266 \pm 0.002$, which exceeds the CF bound by more than $120$ standard deviations (see Fig.~\ref{fig:expsetup}\subref{fig:violationplot}). The uncertainty is evaluated as the standard error of the mean over $100$ repetitions of the measurement, which automatically captures both Poissonian counting statistics and possible additional fluctuations due to experimental instabilities.

Furthermore, we engineer a noisier distribution obtained by collecting the data with an enlarged coincidence window, causing an increase in accidental counts
and acting substantially as a white noise on the correlations. This regime is particularly significant: although the resulting correlations remain nonclassical—violating an instrumental CCM facet—they lie entirely within the CF polytope. 
We observe a violation of a CCM facet by more than $39$ standard deviations, while all $48$ CF facets are satisfied within at least $13$ standard deviations. This provides direct experimental evidence for the existence of correlations that are incompatible with classical causality but compatible with CF.
The locations of both the experimental points in a two-dimensional projection of the OPT, CF, quantum, and CCM correlation sets are shown in Fig.~\ref{fig:results}.

Using the framework introduced in Ref.~\cite{moreno2022events}, we perform an additional analysis where the constraint in Eq.~\eqref{eq:aoe2} is relaxed by allowing non-deterministic outcomes even for the setting $X=0$. Specifically, we can introduce an $\epsilon$-relaxation of Eq.~\eqref{eq:aoe2} with the weaker condition $P(a=c|x=0,c)\geq 1-\epsilon \;\; \forall \,c$. At first sight, the $\epsilon$-relaxation could be interpreted as a trivial tracking failure of the friend, where the friend's record of the variable $C$ is noisy in a classical sense. However, as discussed in Sec.~\ref{sec:assumptions}, such classical noise alone does not affect the CF model. Rather,  we interpret this relaxation as allowing the most general model, compatible with the $\epsilon$-relaxed condition, for reproducing the observed correlations.
With respect to the OPT causal model in Fig.~\ref{fig:instrumental}(c), the $\epsilon$-relaxation corresponds to permitting the nonclassical channel (the direct nonclassical arrow from the nonclassical source to node $A$) to remain active even when $X=0$ in some fraction of the trials, rather than being completely switched off by the friend's classical channel. By characterizing the corresponding enlarged polytope, we find that the observed violation of Eq.~\eqref{eq:genuinelf} in our experiment also excludes models with an $\epsilon$-relaxation up to $\epsilon=13.1\%$ compatible with the observed violation within two standard deviations (see Appendix~\ref{app:relaxationofAOE} for details).

\section{Discussion}

In the last few decades, extensions of Bell’s theorem in two distinct directions have reshaped our understanding of the relation between quantum phenomena and causality.
One extends quantum violations of classical causal models beyond Bell scenarios to networks~\cite{fritz2012BellTheoremCorrelation,henson2014theory,tavakoli2022bell}, including those where communication is allowed and spacelike separation is unnecessary  or impossible,  such as in the instrumental scenario~\cite{chaves2018quantum}. The other, developed through extended Wigner’s friend scenarios, shows that quantum predictions can remain incompatible even with generalized nonclassical causal explanations when Absoluteness of Observed Events is retained~\cite{bong2020strong,cavalcanti2021implications,ying2024relating}, but has so far been restricted to Bell causal structures, with a single source connecting two or more distant parties. 
In this work, we combined these two lines of research, revealing that this stronger form of quantum nonclassicality persists in causal networks beyond the Bell scenario.

We introduced Causal Friendliness in a Wigner’s friend extension of the instrumental network, and used a generalized nonclassical causal framework to study this scenario. We proved that the corresponding quantum correlations are incompatible with any model that jointly satisfies AOE and the generalized OPT causal constraints associated with the instrumental structure. The significance of this result is that it extends the reach of Local-Friendliness-type no-go theorems and the consequent nonclassicality to causal networks beyond Bell scenarios. In particular, it shows that the tension between quantum predictions and the assumptions of AOE and generalized causality appears in scenarios where not only relativistic causal constraints impose no restriction, but which indeed cannot be realized with spacelike separation   (which can be an intricate notion in contexts where AOE is under scrutiny~\cite{cavalcanti2021view}). 

We further realized a proof-of-principle photonic experiment, in which the friend is encoded in the path degree of freedom of a single photon. The observed violation of the corresponding genuine CF inequality is consistent with the quantum prediction and excludes models that preserve both AOE and generalized causality applied to the instrumental Wigner's causal structure. Although the present implementation does not close the detection loophole, it establishes the experimental accessibility of stronger forms of nonclassicality in causal structures beyond Bell.

Beyond the Wigner’s friend framework, our results contribute to the broader program of relaxing the assumptions underlying nonclassicality tests in networks. In a general, theory-independent manner, they show that quantum predictions violate assumptions weaker than those required by classical causal models, even in scenarios without spacelike separation. In this sense, the assumptions about reality challenged by quantum correlations are narrowed down (see Ref.~\cite{vieira2025test} for a parallel approach relaxing standard Bell's theorem assumptions). 
More generally, this work contributes to the broader context of causal networks featuring hybrid OPT-classical assumptions, such as those involving independent sources ~\cite{pozas2022full,wang2023certification,ciudad2024escaping,polino2026network} or broadcasting scenarios~\cite{bowles2020single,villegas2024nonlocality, rodari2023characterizing}. Beyond their foundational implications, our results also point towards new applications for quantum technologies that harness violations of CF in general causal structures, particularly in light of the role of LF-violating correlations in theory-independent randomness certification~\cite{woodhead2014imperfections}.

From a foundational perspective, the violation of CF can only be accommodated by abandoning at least one of the assumptions entering its definition: AOE or one of the two generalized causal constraints associated with the instrumental Wigner's network. 
Interpretations that deny a single observer-independent account of all observed events, including relational~\cite{rovelli1996relational}, many-worlds~\cite{vaidman2024many}, QBist~\cite{fuchs2019qbism}, and other broadly neo-Copenhagen approaches, would explain the violation by rejecting AOE~\cite{schmid2025copenhagenish}. 
By contrast, approaches that retain AOE must be structurally radical, in the sense that they reject the assumed (generalized or classical) causal structure. For instance, de Broglie-Bohm pilot-wave theory~\cite{bohm2006undivided,durr2009bohmian} allows a direct causal influence from Alice's choice to Bob's outcome. Other approaches reject the mathematical constraint that we have called free choice, Eq.~\eqref{eq:frewill}, allowing correlations between Alice's choice and Charlie's observed result, either due to retrocausality \cite{wharton2020colloquium} or superdeterminism~\cite{hossenfelder2020rethinking}.
Finally, spontaneous-collapse models~\cite{bassi2013models} may offer a different route if their dynamics prevent the reversible unitary control required by the protocol, although whether they do so depends on the specific collapse model, its parameters, and the physical implementation of the friend~\cite{wiseman2023thoughtful}.

\section*{Acknowledgments}
The authors thank Joshua Laux for useful discussions and comments. MH thanks Ryan Savage for useful discussions concerning the generalized Bayesian networks framework of Ref.~\cite{henson2014theory}. MH acknowledges use of Claude [Sonnet 5], Large Language Model, (Anthropic, 2026, https://www.anthropic.com) for helping with algebra, and suggesting the solution in Eq.~\eqref{equation:claudeSolution} which was verified by the author.  This work was supported by Australian Research Council (ARC) Discovery Early Career Researcher Award (DE250100762) (E.P.) and ARC Discovery Project DP25010216 (H.M.W, N.T, E.C). A.D. acknowledges
support by the Australian Government Research Training Program (RTP). The authors acknowledge support from FARE Ricerca in Italia QU-DICE Grant n. R20TRHTSPA and by the ERC Advanced Grant QU-BOSS (QUantum advantage via nonlinear BOSon Sampling, grant agreement no. 884676).

\nocite{mukherjee2026limits}
\bibliography{biblionew.bib}

@article{brunner2014bell,
    author = "Brunner, Nicolas and Cavalcanti, Daniel and Pironio, Stefano and Scarani, Valerio and Wehner, Stephanie",
    title = "{Bell nonlocality}",
    eprint = "1303.2849",
    archivePrefix = "arXiv",
    primaryClass = "quant-ph",
    doi = "10.1103/RevModPhys.86.419",
    journal = "Rev. Mod. Phys.",
    volume = "86",
    pages = "419",
    year = "2014"
}

@article{hossenfelder2020rethinking,
    author = "Hossenfelder, S. and Palmer, T. N.",
    title = "{Rethinking Superdeterminism}",
    eprint = "1912.06462",
    archivePrefix = "arXiv",
    primaryClass = "quant-ph",
    doi = "10.3389/fphy.2020.00139",
    journal = "Front. in Phys.",
    volume = "8",
    pages = "139",
    year = "2020"
}

@article{allen2017quantum,
    author = "Allen, John-Mark A. and Barrett, Jonathan and Horsman, Dominic C. and Lee, Ciar{\'a}n M. and Spekkens, Robert W.",
    title = "{Quantum Common Causes and Quantum Causal Models}",
    eprint = "1609.09487",
    archivePrefix = "arXiv",
    primaryClass = "quant-ph",
    doi = "10.1103/PhysRevX.7.031021",
    journal = "Phys. Rev. X",
    volume = "7",
    number = "3",
    pages = "031021",
    year = "2017"
}

@article{costa2016quantum,
    author = "Costa, Fabio and Shrapnel, Sally",
    title = "{Quantum causal modelling}",
    eprint = "1512.07106",
    archivePrefix = "arXiv",
    primaryClass = "quant-ph",
    doi = "10.1088/1367-2630/18/6/063032",
    journal = "New J. Phys.",
    volume = "18",
    number = "6",
    pages = "063032",
    year = "2016"
}

@misc{barrett2019quantum,
    author = "Barrett, Jonathan and Lorenz, Robin and Oreshkov, Ognyan",
    title = "{Quantum Causal Models}",
    eprint = "1906.10726",
    archivePrefix = "arXiv",
    primaryClass = "quant-ph",
    month = "6",
    year = "2019"
}

@article{wolfe2021quantum,
    author = "Wolfe, Elie and Pozas-Kerstjens, Alejandro and Grinberg, Matan and Rosset, Denis and Ac{\'\i}n, Antonio and Navascues, Miguel",
    title = "{Quantum Inflation: A General Approach to Quantum Causal Compatibility}",
    eprint = "1909.10519",
    archivePrefix = "arXiv",
    primaryClass = "quant-ph",
    doi = "10.1103/PhysRevX.11.021043",
    journal = "Phys. Rev. X",
    volume = "11",
    number = "2",
    pages = "021043",
    year = "2021"
}

@article{fedrizzi2007wavelength,
    author = "Fedrizzi, Alessandro and Herbst, Thomas and Poppe, Andreas and Jennewein, Thomas and Zeilinger, Anton",
    title = "{A wavelength-tunable fiber-coupled source of narrowband entangled photons}",
    eprint = "0706.2877",
    archivePrefix = "arXiv",
    primaryClass = "quant-ph",
    doi = "10.1364/OE.15.015377",
    journal = "Opt. Express",
    volume = "15",
    number = "23",
    pages = "15377--15386",
    year = "2007"
}

@article{ying2024relating,
    author = "Y{\={i}}ng, Y{\`\i}l{\`e} and Ansanelli, Marina Maciel and Di Biagio, Andrea and Wolfe, Elie and Schmid, David and Cavalcanti, Eric Gama",
    title = "{Relating Wigner's Friend Scenarios to Nonclassical Causal Compatibility, Monogamy Relations, and Fine Tuning}",
    eprint = "2309.12987",
    archivePrefix = "arXiv",
    primaryClass = "quant-ph",
    doi = "10.22331/q-2024-09-26-1485",
    journal = "Quantum",
    volume = "8",
    pages = "1485",
    year = "2024"
}

@article{van2019quantum,
    author = "Van Himbeeck, Thomas and Brask, Jonatan Bohr and Pironio, Stefano and Ramanathan, Ravishankar and Sainz, Ana Bel{\'e}n and Wolfe, Elie",
    title = "{Quantum violations in the Instrumental scenario and their relations to the Bell scenario}",
    eprint = "1804.04119",
    archivePrefix = "arXiv",
    primaryClass = "quant-ph",
    doi = "10.22331/q-2019-09-16-186",
    journal = "Quantum",
    volume = "3",
    pages = "186",
    year = "2019"
}

@inproceedings{pearl1995testability,
  title={On the testability of causal models with latent and instrumental variables},
  author={Pearl, Judea},
  booktitle={Proceedings of the Eleventh conference on Uncertainty in artificial intelligence},
  pages={435--443},
  year={1995},
  eprint={1302.4976},
  archivePrefix={arXiv},
  primaryClass={cs.AI},
  url={https://arxiv.org/abs/1302.4976}, 
}

@article{chiribella2010probabilistic,
    author = "Chiribella, Giulio and D{\textquoteright}Ariano, Giacomo Mauro and Perinotti, Paolo",
    title = "{Probabilistic theories with purification}",
    eprint = "0908.1583",
    archivePrefix = "arXiv",
    primaryClass = "quant-ph",
    doi = "10.1103/PhysRevA.81.062348",
    journal = "Phys. Rev. A",
    volume = "81",
    number = "6",
    pages = "062348",
    year = "2010"
}

@article{cabello2013basic,
    author = "Cabello, Adan and Danielsen, Lars Eirik and Lopez-Tarrida, Antonio J. and Portillo, Jose R.",
    title = "{Basic exclusivity graphs in quantum correlations}",
    eprint = "1211.5825",
    archivePrefix = "arXiv",
    primaryClass = "quant-ph",
    doi = "10.1103/PhysRevA.88.032104",
    journal = "Phys. Rev. A",
    volume = "88",
    pages = "032104",
    year = "2013"
}

@article{cabello2013simple,
    author = "Cabello, Adan",
    title = "{Simple explanation of the quantum violation of a fundamental inequality}",
    eprint = "1210.2988",
    archivePrefix = "arXiv",
    primaryClass = "quant-ph",
    doi = "10.1103/PhysRevLett.110.060402",
    journal = "Phys. Rev. Lett.",
    volume = "110",
    pages = "060402",
    year = "2013"
}

@article{proietti2019experimental,
    author = "Proietti, Massimiliano and Pickston, Alexander and Graffitti, Francesco and Barrow, Peter and Kundys, Dmytro and Branciard, Cyril and Ringbauer, Martin and Fedrizzi, Alessandro",
    title = "{Experimental test of local observer-independence}",
    journal = {Science Advances},
    volume = {5},
    number = {9},
    pages = {eaaw9832},
    year = {2019},
    doi = {10.1126/sciadv.aaw9832},
    URL = {https://www.science.org/doi/abs/10.1126/sciadv.aaw9832},
    eprint = "1902.05080",
    archivePrefix = "arXiv",
    primaryClass = "quant-ph",
    month = "11",
    year = "2019"
}

@article{fine1970insolubility,
  title = {Insolubility of the Quantum Measurement Problem},
  author = {Fine, Arthur},
  journal = {Phys. Rev. D},
  volume = {2},
  issue = {12},
  pages = {2783--2787},
  numpages = {0},
  year = {1970},
  month = {Dec},
  publisher = {American Physical Society},
  doi = {10.1103/PhysRevD.2.2783},
  url = {https://link.aps.org/doi/10.1103/PhysRevD.2.2783}
}

@inbook{brukner2017quantum,
author="Brukner, {\v{C}}aslav",
editor="Bertlmann, Reinhold
and Zeilinger, Anton",
title="On the Quantum Measurement Problem",
bookTitle="Quantum [Un]Speakables II: Half a Century of Bell's Theorem",
year="2017",
publisher="Springer International Publishing",
address="Cham",
pages="95--117",
isbn="978-3-319-38987-5",
doi="10.1007/978-3-319-38987-5_5",
url="https://doi.org/10.1007/978-3-319-38987-5_5",
eprint = "1507.05255",
archivePrefix = "arXiv",
primaryClass = "quant-ph",
}

@article{pozas2022full,
    author = "Pozas-Kerstjens, Alejandro and Gisin, Nicolas and Tavakoli, Armin",
    title = "{Full Network Nonlocality}",
    eprint = "2105.09325",
    archivePrefix = "arXiv",
    primaryClass = "quant-ph",
    doi = "10.1103/PhysRevLett.128.010403",
    journal = "Phys. Rev. Lett.",
    volume = "128",
    number = "1",
    pages = "010403",
    year = "2022"
}

@article{cavalcanti2021implications,
    author = "Cavalcanti, Eric G. and Wiseman, Howard M.",
    title = "{Implications of Local Friendliness violation for quantum causality}",
    eprint = "2106.04065",
    archivePrefix = "arXiv",
    primaryClass = "quant-ph",
    doi = "10.3390/e23080925",
    journal = "Entropy",
    volume = "23",
    pages = "925",
    year = "2021"
}

@article{ormrod2026theories,
    author = "Ormrod, Nick and Vilasini, V. and Barrett, Jonathan",
    title = "{Which theories have a measurement problem?}",
    eprint = "2303.03353",
    archivePrefix = "arXiv",
    primaryClass = "quant-ph",
    doi = "10.1088/1367-2630/ae131e",
    journal = "New J. Phys.",
    volume = "28",
    number = "2",
    pages = "024506",
    year = "2026"
}

@article{vieira2025test,
    author = "Vieira, Carlos and Ramanathan, Ravishankar and Cabello, Ad{\'a}n",
    title = "{Test of the physical significance of Bell non-locality}",
    eprint = "2402.00801",
    archivePrefix = "arXiv",
    primaryClass = "quant-ph",
    doi = "10.1038/s41467-025-59247-7",
    journal = "Nature Commun.",
    volume = "16",
    number = "1",
    pages = "4390",
    year = "2025"
}

@incollection{wigner1995remarks,
author="Wigner, E. P.",
editor="Mehra, Jagdish",
title="Remarks on the Mind-Body Question",
bookTitle="Philosophical Reflections and Syntheses",
year="1995",
publisher="Springer Berlin Heidelberg",
address="Berlin, Heidelberg",
pages="247--260",
isbn="978-3-642-78374-6",
doi="10.1007/978-3-642-78374-6_20",
url="https://doi.org/10.1007/978-3-642-78374-6_20"
}

@article{deutsch1985quantum,
    author = "Deutsch, David",
    title = "{Quantum theory as a universal physical theory}",
    doi = "10.1007/BF00670071",
    journal = "Int. J. Theor. Phys.",
    volume = "24",
    number = "1",
    pages = "1--41",
    year = "1985"
}

@article{wigner1984review,
  title={Review of the quantum-mechanical measurement problem},
  author={Wigner, Eugene P},
  journal={Science, Computers, and the Information Onslaught},
  pages={63--82},
  year={1984},
  publisher={Elsevier}
}

@misc{gisin2018collapse,
    author = "Gisin, Nicolas",
    title = "{Collapse. What else?}",
    eprint = "1701.08300",
    archivePrefix = "arXiv",
    primaryClass = "quant-ph",
    reportNumber = "in Collapse of the Wave Function, Ed. Shan Gao, Cambridge Uni Press, pp 207-224 (2018)",
    month = "4",
    year = "2017"
}

@article{bell1990against,
  title={Against ‘measurement’},
  author={Bell, John},
  journal={Physics world},
  volume={3},
  number={8},
  pages={33},
  year={1990},
  publisher={IOP Publishing}
}

@misc{bub2010two,
    author = "Bub, Jeffrey and Pitowsky, Itamar",
    title = "{Two dogmas about quantum mechanics}",
    eprint = "0712.4258",
    archivePrefix = "arXiv",
    primaryClass = "quant-ph",
    month = "5",
    year = "2008"
}

@article{maudlin1995three,
  author    = {Maudlin, Tim},
  title     = {Three measurement problems},
  journal   = {Topoi},
  year      = {1995},
  month     = mar,
  volume    = {14},
  number    = {1},
  pages     = {7--15},
  issn      = {1572-8749},
  doi       = {10.1007/BF00763473},
  url       = {https://doi.org/10.1007/BF00763473}
}

@incollection{wigner1997problem,
author="Wigner, E. P.",
editor="Mehra, Jagdish",
title="The Problem of Measurement",
bookTitle="Philosophical Reflections and Syntheses",
year="1995",
publisher="Springer Berlin Heidelberg",
address="Berlin, Heidelberg",
pages="163--180",
isbn="978-3-642-78374-6",
doi="10.1007/978-3-642-78374-6_12",
url="https://doi.org/10.1007/978-3-642-78374-6_12"
}

@misc{schmid2025copenhagenish,
    author = "Schmid, David and Y{\={i}}ng, Y{\`\i}l{\`e} and Leifer, Matthew",
    title = "{Copenhagenish interpretations of quantum mechanics}",
    eprint = "2506.00112",
    archivePrefix = "arXiv",
    primaryClass = "quant-ph",
    month = "5",
    year = "2025"
}

@misc{schmid2023review,
    author = "Schmid, David and Y{\={i}}ng, Y{\`\i}l{\`e} and Leifer, Matthew",
    title = "{A review and analysis of six extended Wigner's friend arguments}",
    eprint = "2308.16220",
    archivePrefix = "arXiv",
    primaryClass = "quant-ph",
    month = "8",
    year = "2023"
}

@article{cavalcanti2021view,
    author = "Cavalcanti, Eric G.",
    title = "{The view from a Wigner bubble}",
    eprint = "2008.05100",
    archivePrefix = "arXiv",
    primaryClass = "quant-ph",
    doi = "10.1007/s10701-021-00417-0",
    journal = "Found. Phys.",
    volume = "51",
    pages = "39",
    year = "2021"
}

@misc{cavalcanti2016bell,
    author = "Cavalcanti, Eric G.",
    title = "{Bell's theorem and the measurement problem: reducing two mysteries to one?}",
    eprint = "1602.07404",
    archivePrefix = "arXiv",
    primaryClass = "quant-ph",
    doi = "10.1088/1742-6596/701/1/012002",
    month = "2",
    year = "2016"
}

@inproceedings{wiseman2017causarum,
    author = "Wiseman, Howard M. and Cavalcanti, Eric G.",
    title = "{Causarum Investigatio and the Two Bell's Theorems of John Bell}",
    booktitle = "{Quantum [Un]Speakables II: 50 Years of Bell's Theorem}: {A celebration of John Bell, his theorem and its consequences}",
    eprint = "1503.06413",
    archivePrefix = "arXiv",
    primaryClass = "quant-ph",
    month = "3",
    year = "2015"
}

@misc{haddara2025properties,
    author = "Haddara, Marwan and Wiseman, Howard M. and Cavalcanti, Eric G.",
    title = "{Properties and Applications of Partially Deterministic Polytopes}",
    eprint = "2510.25127",
    archivePrefix = "arXiv",
    primaryClass = "quant-ph",
    month = "10",
    year = "2025"
}

@article{adlam2024does,
    author = "Adlam, Emily",
    title = "{What Does {\textquoteleft}(Non)-absoluteness of Observed Events{\textquoteright} Mean?}",
    eprint = "2309.03171",
    archivePrefix = "arXiv",
    primaryClass = "quant-ph",
    doi = "10.1007/s10701-023-00747-1",
    journal = "Found. Phys.",
    volume = "54",
    number = "1",
    pages = "13",
    year = "2024"
}

@article{chaves2021causal,
    author = {Chaves, Rafael and Moreno, George and Polino, Emanuele and Poderini, Davide and Agresti, Iris and Suprano, Alessia and Barros, Mariana R. and Carvacho, Gonzalo and Wolfe, Elie and Canabarro, Askery and Spekkens, Robert W. and Sciarrino, Fabio},
    title = "{Causal Networks and Freedom of Choice in Bell{\textquoteright}s Theorem}",
    eprint = "2105.05721",
    archivePrefix = "arXiv",
    primaryClass = "quant-ph",
    doi = "10.1103/PRXQuantum.2.040323",
    journal = "PRX Quantum",
    volume = "2",
    number = "4",
    pages = "040323",
    year = "2021"
}

@article{moreno2022events,
    author = "Moreno, George and Nery, Ranieri and Duarte, Cristhiano and Chaves, Rafael",
    title = "{Events in quantum mechanics are maximally non-absolute}",
    eprint = "2112.11223",
    archivePrefix = "arXiv",
    primaryClass = "quant-ph",
    doi = "10.22331/q-2022-08-24-785",
    journal = "Quantum",
    volume = "6",
    pages = "785",
    year = "2022"
}

@article{wiseman2023thoughtful,
    author = "Wiseman, Howard M. and Cavalcanti, Eric G. and Rieffel, Eleanor G.",
    title = "{A ''thoughtful'' Local Friendliness no-go theorem: a prospective experiment with new assumptions to suit}",
    eprint = "2209.08491",
    archivePrefix = "arXiv",
    primaryClass = "quant-ph",
    doi = "10.22331/q-2023-09-14-1112",
    journal = "Quantum",
    volume = "7",
    pages = "1112",
    year = "2023"
}

@book{durr2009bohmian, title={Bohmian mechanics. The physics and mathematics of quantum theory}, DOI={10.1007/b99978}, abstractNote={Bohmian Mechanics was formulated in 1952 by David Bohm as a complete theory of quantum phenomena based on a particle picture. It was promoted some decades later by John S. Bell, who, intrigued by the manifestly nonlocal structure of the theory, was led to his famous Bell's inequalities. Experimental tests of the inequalities verified that nature is indeed nonlocal. Bohmian mechanics has since then prospered as the straightforward completion of quantum mechanics. This book provides a systematic introduction to Bohmian mechanics and to the mathematical abstractions of quantum mechanics, which range from the self-adjointness of the Schroedinger operator to scattering theory. It explains how the quantum formalism emerges when Boltzmann's ideas about statistical mechanics are applied to Bohmian mechanics. The book is self-contained, mathematically rigorous and an ideal starting point for a fundamental approach to quantum mechanics. It will appeal to students and newcomers to the field, as well as to established scientists seeking a clear exposition of the theory. (orig.)}, publisher={Springer}, author={Duerr, Detlef and Teufel, Stefan}, year={2009},
}

@article{chaves2015unifying,
    author = "Chaves, Rafael and Kueng, Richard and Brask, Jonatan Bohr and Gross, David",
    title = "{A unifying framework for relaxations of the causal assumptions in Bell's theorem}",
    eprint = "1411.4648",
    archivePrefix = "arXiv",
    primaryClass = "quant-ph",
    doi = "10.1103/PhysRevLett.114.140403",
    journal = "Phys. Rev. Lett.",
    volume = "114",
    pages = "140403",
    year = "2015"
}

@article{wharton2020colloquium,
    author = "Wharton, K. B. and Argaman, N.",
    title = "{Colloquium : Bell{\textquoteright}s theorem and locally mediated reformulations of quantum mechanics}",
    eprint = "1906.04313",
    archivePrefix = "arXiv",
    primaryClass = "quant-ph",
    doi = "10.1103/RevModPhys.92.021002",
    journal = "Rev. Mod. Phys.",
    volume = "92",
    number = "2",
    pages = "021002",
    year = "2020"
}

@book{bohm2006undivided,
	author = {David Bohm and Basil J. Hiley},
	editor = {},
	publisher = {Routledge},
	title = {The Undivided Universe: An Ontological Interpretation of Quantum Theory},
	year = {1995}
}

@article{bassi2013models,
    author = "Bassi, Angelo and Lochan, Kinjalk and Satin, Seema and Singh, Tejinder P. and Ulbricht, Hendrik",
    title = "{Models of Wave-function Collapse, Underlying Theories, and Experimental Tests}",
    eprint = "1204.4325",
    archivePrefix = "arXiv",
    primaryClass = "quant-ph",
    doi = "10.1103/RevModPhys.85.471",
    journal = "Rev. Mod. Phys.",
    volume = "85",
    pages = "471--527",
    year = "2013"
}

@misc{vaidman2024many,
  title={The many-worlds interpretation of quantum mechanics: Current status and relation to other interpretations},
  author={Vaidman, Lev},
  journal={Quantum Reports},
  volume={6},
  number={2},
  pages={142--146},
  year={2024},
  publisher={MDPI}
}

@article{fuchs2019qbism,
    author = "Fuchs, Christopher A. and Stacey, Blake C.",
    title = "{QBism: Quantum theory as a hero's handbook}",
    doi = "10.3254/978-1-61499-937-9-133",
    journal = "Proc. Int. Sch. Phys. Fermi",
    volume = "197",
    pages = "133--202",
    year = "2019"
}

@article{rovelli1996relational,
  author    = {Rovelli, Carlo},
  title     = {Relational quantum mechanics},
  journal   = {International Journal of Theoretical Physics},
  year      = {1996},
  month     = aug,
  volume    = {35},
  number    = {8},
  pages     = {1637--1678},
  issn      = {1572-9575},
  doi       = {10.1007/BF02302261},
  url       = {https://doi.org/10.1007/BF02302261}
}

@article{ciudad2024escaping,
    author = "Ala{\~n}{\'o}n, Maria Ciudad and Boghiu, Emanuel-Cristian and Abiuso, Paolo and Wolfe, Elie",
    title = "{Escaping the Shadow of Bell's Theorem in Network Nonlocality}",
    eprint = "2406.15587",
    archivePrefix = "arXiv",
    primaryClass = "quant-ph",
    doi = "10.22331/q-2026-09-21-2213",
    journal = "Quantum",
    volume = "10",
    pages = "2213",
    year = "2026"
}

@article{cavalcanti2014modifications,
    author = "Cavalcanti, Eric G. and Lal, Raymond",
    title = "{On modifications of Reichenbach's principle of common cause in light of Bell's theorem}",
    eprint = "1311.6852",
    archivePrefix = "arXiv",
    primaryClass = "quant-ph",
    doi = "10.1088/1751-8113/47/42/424018",
    journal = "J. Phys. A",
    volume = "47",
    pages = "424018",
    year = "2014"
}

@article{wang2026experimental,
  author={Wang, Ning-Ning and Zhang, Chao and Cao, Huan and Xu, Kai and Liu, Bi-Heng and Huang, Yun-Feng and Li, Chuan-Feng and Guo, Guang-Can and Gisin, Nicolas and Kriv{\'a}chy, Tam{\'a}s and others},    title = "{Experimental Genuine Quantum Nonlocality in the Triangle Network}",
  eprint = "2401.15428",
  archivePrefix = "arXiv",
  primaryClass = "quant-ph",
  doi = "10.1103/5hhc-rw3t",
  journal = "Phys. Rev. Lett.",
  volume = "136",
  number = "18",
  pages = "180202",
  year = "2026"
}

@article{renou2019genuine,
    author = {Renou, Marc-Olivier and B{\"a}umer, Elisa and Boreiri, Sadra and Brunner, Nicolas and Gisin, Nicolas and Beigi, Salman},
    title = "{Genuine Quantum Nonlocality in the Triangle Network}",
    eprint = "1905.04902",
    archivePrefix = "arXiv",
    primaryClass = "quant-ph",
    doi = "10.1103/PhysRevLett.123.140401",
    journal = "Phys. Rev. Lett.",
    volume = "123",
    number = "14",
    pages = "140401",
    year = "2019"
}

@article{branciard2012bilocal,
    author = "Branciard, Cyril and Rosset, Denis and Gisin, Nicolas and Pironio, Stefano",
    title = "{Bilocal versus nonbilocal correlations in entanglement-swapping experiments}",
    eprint = "1112.4502",
    archivePrefix = "arXiv",
    primaryClass = "quant-ph",
    doi = "10.1103/PhysRevA.85.032119",
    journal = "Phys. Rev. A",
    volume = "85",
    number = "3",
    pages = "032119",
    year = "2012"
}

@article{balke1997bounds,
    author = {Alexander Balke and Judea Pearl},
    title = {Bounds on Treatment Effects from Studies with Imperfect Compliance},
    journal = {Journal of the American Statistical Association},
    volume = {92},
    number = {439},
    pages = {1171--1176},
    year = {1997},
    publisher = {Taylor \& Francis},
    doi = {10.1080/01621459.1997.10474074},
    URL = {https://doi.org/10.1080/01621459.1997.10474074},
    eprint = {https://doi.org/10.1080/01621459.1997.10474074}
}

@article{tavakoli2022bell,
    author = "Tavakoli, Armin and Pozas-Kerstjens, Alejandro and Luo, Ming-Xing and Renou, Marc-Olivier",
    title = "{Bell nonlocality in networks}",
    eprint = "2104.10700",
    archivePrefix = "arXiv",
    primaryClass = "quant-ph",
    doi = "10.1088/1361-6633/ac41bb",
    journal = "Rept. Prog. Phys.",
    volume = "85",
    number = "5",
    pages = "056001",
    year = "2022"
}

@article{polino2023experimental,
  author={Polino, Emanuele and Poderini, Davide and Rodari, Giovanni and Agresti, Iris and Suprano, Alessia and Carvacho, Gonzalo and Wolfe, Elie and Canabarro, Askery and Moreno, George and Milani, Giorgio and Spekkens, Robert W. and Chaves, Rafael and Sciarrino, Fabio},
  title = "{Experimental nonclassicality in a causal network without assuming freedom of choice}",
  eprint = "2210.07263",
  archivePrefix = "arXiv",
  primaryClass = "quant-ph",
  doi = "10.1038/s41467-023-36428-w",
  journal = "Nature Commun.",
  volume = "14",
  number = "1",
  pages = "909",
  year = "2023"
}

@article{haddara2025local,
    author = "Haddara, Marwan and Cavalcanti, Eric G.",
    title = "{Local friendliness polytopes in multipartite scenarios}",
    eprint = "2407.20346",
    archivePrefix = "arXiv",
    primaryClass = "quant-ph",
    doi = "10.1103/PhysRevA.111.012206",
    journal = "Phys. Rev. A",
    volume = "111",
    number = "1",
    pages = "012206",
    year = "2025"
}

@article{brukner2018no,
    author = "Brukner, Caslav",
    title = "{A no-go theorem for observer-independent facts}",
    eprint = "1804.00749",
    archivePrefix = "arXiv",
    primaryClass = "quant-ph",
    doi = "10.3390/e20050350",
    journal = "Entropy",
    volume = "20",
    pages = "350",
    year = "2018"
}

@article{bong2020strong,
    author = "Bong, Kok-Wei and Utreras-Alarc{\'o}n, An{\'\i}bal and Ghafari, Farzad and Liang, Yeong-Cherng and Tischler, Nora and Cavalcanti, Eric G. and Pryde, Geoff J. and Wiseman, Howard M.",
    title = "{A strong no-go theorem on the Wigner{\textquoteright}s friend paradox}",
    eprint = "1907.05607",
    archivePrefix = "arXiv",
    primaryClass = "quant-ph",
    doi = "10.1038/s41567-020-0990-x",
    journal = "Nature Phys.",
    volume = "16",
    number = "12",
    pages = "1199--1205",
    year = "2020"
}

@article{brukner2022wigner,
  author    = {Brukner, {\v{C}}aslav},
  title     = {Wigner's friend and relational objectivity},
  journal   = {Nature Reviews Physics},
  year      = {2022},
  month     = oct,
  volume    = {4},
  number    = {10},
  pages     = {628--630},
  issn      = {2522-5820},
  doi       = {10.1038/s42254-022-00505-8},
  url       = {https://doi.org/10.1038/s42254-022-00505-8}
}

@article{frauchiger2018quantum,
    author = "Frauchiger, Daniela and Renner, Renato",
    title = "{Quantum theory cannot consistently describe the use of itself}",
    eprint = "1604.07422",
    archivePrefix = "arXiv",
    primaryClass = "quant-ph",
    doi = "10.1038/s41467-018-05739-8",
    journal = "Nature Commun.",
    volume = "9",
    number = "1",
    pages = "3711",
    year = "2018"
}

@misc{rodari2023characterizing,
    author = "Rodari, Giovanni and Poderini, Davide and Polino, Emanuele and Suprano, Alessia and Sciarrino, Fabio and Chaves, Rafael",
    title = "{Characterizing Hybrid Causal Structures with the Exclusivity Graph Approach}",
    eprint = "2401.00063",
    archivePrefix = "arXiv",
    primaryClass = "quant-ph",
    month = "12",
    year = "2023"
}

@article{fritz2012BellTheoremCorrelation,
  title = {Beyond {{Bell}}'s Theorem: Correlation Scenarios},
  shorttitle = {Beyond {{Bell}}'s Theorem},
  author = {Fritz, Tobias},
  year = {2012},
  month = oct,
  journal = {New Journal of Physics},
  volume = {14},
  number = {10},
  pages = {103001},
  publisher = {{IOP Publishing}},
  issn = {1367-2630},
  doi = {2012100112130300},
  urldate = {2023-07-10},
  langid = {english}
}

@article{gu2023experimental,
  author={Gu, Xue-Mei and Huang, Liang and Pozas-Kerstjens, Alejandro and Jiang, Yang-Fan and Wu, Dian and Bai, Bing and Sun, Qi-Chao and Chen, Ming-Cheng and Zhang, Jun and Yu, Sixia and Zhang, Qiang and Lu, Chao-Yang and Pan, Jian-Wei},
  title = "{Experimental Full Network Nonlocality with Independent Sources and Strict Locality Constraints}",
  eprint = "2302.02472",
  archivePrefix = "arXiv",
  primaryClass = "quant-ph",
  doi = "10.1103/PhysRevLett.130.190201",
  journal = "Phys. Rev. Lett.",
  volume = "130",
  number = "19",
  pages = "190201",
  year = "2023"
}

@article{chaves2018quantum,
    author = "Chaves, Rafael and Carvacho, Gonzalo and Agresti, Iris and Di Giulio, Valerio and Aolita, Leandro and Giacomini, Sandro and Sciarrino, Fabio",
    title = "{Quantum violation of an instrumental test}",
    eprint = "1808.09894",
    archivePrefix = "arXiv",
    primaryClass = "quant-ph",
    doi = "10.1038/s41567-017-0008-5",
    journal = "Nature Phys.",
    volume = "14",
    pages = "291",
    year = "2018"
}

@article{poderini2020experimental,
  title={Experimental violation of n-locality in a star quantum network},
  author={Poderini, Davide and Agresti, Iris and Marchese, Guglielmo and Polino, Emanuele and Giordani, Taira and Suprano, Alessia and Valeri, Mauro and Milani, Giorgio and Spagnolo, Nicol{\`o} and Carvacho, Gonzalo and Chaves, Rafael and Sciarrino, Fabio},
  journal={Nature Communications},
  volume={11},
  number={1},
  pages={1--8},
  year={2020},
  publisher={Nature Publishing Group},
  doi={https://doi.org/10.1038/s41467-020-16189-6}
}

@article{wang2023certification,
    author = "Wang, Ning-Ning and Pozas-Kerstjens, Alejandro and Zhang, Chao and Liu, Bi-Heng and Huang, Yun-Feng and Li, Chuan-Feng and Guo, Guang-Can and Gisin, Nicolas and Tavakoli, Armin",
    title = "{Certification of non-classicality in all links of a photonic star network without assuming quantum mechanics}",
    eprint = "2212.09765",
    archivePrefix = "arXiv",
    primaryClass = "quant-ph",
    doi = "10.1038/s41467-023-37842-w",
    journal = "Nature Commun.",
    volume = "14",
    number = "1",
    pages = "2153",
    year = "2023"
}

@article{saunders2017experimental,
    author = "Saunders, Dylan J. and Bennet, Adam J. and Branciard, Cyril and Pryde, Geoff J.",
    title = "{Experimental demonstration of nonbilocal quantum correlations}",
    eprint = "1610.08514",
    archivePrefix = "arXiv",
    primaryClass = "quant-ph",
    doi = "10.1126/sciadv.1602743",
    journal = "Sci. Adv.",
    volume = "3",
    number = "4",
    pages = "1602743",
    year = "2017"
}

@article{carvacho2017experimental,
    author = "Carvacho, Gonzalo and Andreoli, Francesco and Santodonato, Luca and Bentivegna, Marco and Chaves, Rafael and Sciarrino, Fabio",
    title = "{Experimental violation of local causality in a quantum network}",
    eprint = "1610.03327",
    archivePrefix = "arXiv",
    primaryClass = "quant-ph",
    doi = "10.1038/ncomms14775",
    journal = "Nature Commun.",
    volume = "8",
    number = "1",
    pages = "14775",
    year = "2017"
}

@article{bell1964einstein,
    author = "Bell, J. S.",
    title = "{On the Einstein-Podolsky-Rosen paradox}",
    reportNumber = "RX-1376",
    doi = "10.1103/PhysicsPhysiqueFizika.1.195",
    journal = "Physics Physique Fizika",
    volume = "1",
    pages = "195--200",
    year = "1964"
}

@book{pearl2009causality,
  title={Causality},
  author={Pearl, Judea},
  year={2009},
  publisher={Cambridge university press}
}

@article{mukherjee2026limits,
    author = "Mukherjee, Sumit and Hance, Jonte R.",
    title = "{Limits of absoluteness of observed events in timelike scenarios: A no-go theorem}",
    eprint = "2510.26562",
    archivePrefix = "arXiv",
    primaryClass = "quant-ph",
    doi = "10.1103/1c6t-45g8",
    journal = "Phys. Rev. Res.",
    volume = "8",
    number = "2",
    pages = "023066",
    year = "2026"
}

@article{bowles2020single,
    author = "Bowles, Joseph and Hirsch, Flavien and Cavalcanti, Daniel",
    title = "{Single-copy activation of Bell nonlocality via broadcasting of quantum states}",
    eprint = "2007.16034",
    archivePrefix = "arXiv",
    primaryClass = "quant-ph",
    doi = "10.22331/q-2021-07-13-499",
    journal = "Quantum",
    volume = "5",
    pages = "499",
    year = "2021"
}

@article{cabello2014graph,
    author = "Cabello, Ad{\'a}n and Severini, Simone and Winter, Andreas",
    title = "{Graph-Theoretic Approach to Quantum Correlations}",
    eprint = "1401.7081",
    archivePrefix = "arXiv",
    primaryClass = "quant-ph",
    doi = "10.1103/PhysRevLett.112.040401",
    journal = "Phys. Rev. Lett.",
    volume = "112",
    number = "4",
    pages = "040401",
    year = "2014"
}

@misc{cabello2012specker,
    author = "Cabello, Adan",
    title = "{Specker's fundamental principle of quantum mechanics}",
    eprint = "1212.1756",
    archivePrefix = "arXiv",
    primaryClass = "quant-ph",
    month = "12",
    year = "2012"
}

@article{acin2015combinatorial,
    author = "Ac{\'\i}n, Antonio and Fritz, Tobias and Leverrier, Anthony and Sainz, Ana Bel{\'e}n",
    title = "{A Combinatorial Approach to Nonlocality and Contextuality}",
    eprint = "1212.4084",
    archivePrefix = "arXiv",
    primaryClass = "quant-ph",
    doi = "10.1007/s00220-014-2260-1",
    journal = "Commun. Math. Phys.",
    volume = "334",
    number = "2",
    pages = "533--628",
    year = "2015"
}

@inproceedings{poderini2020exclusivity,
  title={Exclusivity graph approach to instrumental inequalities},
  author={Poderini, Davide and Chaves, Rafael and Agresti, Iris and Carvacho, Gonzalo and Sciarrino, Fabio},
  booktitle={Uncertainty in Artificial Intelligence},
  pages={1274--1283},
  year={2020},
  organization={PMLR},
  eprint = "1909.09120",
  archivePrefix = "arXiv",
  primaryClass = "quant-ph",
  month = "9"
}

@misc{knuth1993sandwich,
  title={The Sandwich Theorem},
  author={Donald Ervin Knuth},
  journal={Electron. J. Comb.},
  year={1993},
  volume={1},
  eprint={math/9312214},
  archivePrefix={arXiv},
  primaryClass={math.CO},
  url={https://arxiv.org/abs/math/9312214} 
}

@article{grotschel1986relaxations,
title = {Relaxations of vertex packing},
journal = {Journal of Combinatorial Theory, Series B},
volume = {40},
number = {3},
pages = {330-343},
year = {1986},
issn = {0095-8956},
doi = {https://doi.org/10.1016/0095-8956(86)90087-0},
url = {https://www.sciencedirect.com/science/article/pii/0095895686900870},
author = {M Grötschel and L Lovász and A Schrijver}
}

@article{henson2014theory,
    author = "Pusey, Matthew F. and Lal, Raymond and Henson, Joe",
    title = "{Theory-independent limits on correlations from generalized Bayesian networks}",
    eprint = "1405.2572",
    archivePrefix = "arXiv",
    primaryClass = "quant-ph",
    doi = "10.1088/1367-2630/16/11/113043",
    journal = "New J. Phys.",
    volume = "16",
    number = "11",
    pages = "113043",
    year = "2014"
}

@article{wood2015lesson,
    author = "Wood, Christopher J. and Spekkens, Robert W.",
    title = "{The lesson of causal discovery algorithms for quantum correlations: causal explanations of Bell-inequality violations require fine-tuning}",
    eprint = "1208.4119",
    archivePrefix = "arXiv",
    primaryClass = "quant-ph",
    doi = "10.1088/1367-2630/17/3/033002",
    journal = "New J. Phys.",
    volume = "17",
    number = "3",
    pages = "033002",
    year = "2015"
}

@article{navascues2008convergent,
    author = "Ac{\'\i}n, Antonio and Pironio, Stefano and Navascu{\'e}s, Miguel",
    title = "{A convergent hierarchy of semidefinite programs characterizing the set of quantum correlations}",
    eprint = "0803.4290",
    archivePrefix = "arXiv",
    primaryClass = "quant-ph",
    doi = "10.1088/1367-2630/10/7/073013",
    journal = "New J. Phys.",
    volume = "10",
    number = "7",
    pages = "073013",
    year = "2008"
}

@article{agresti2022experimental,
    author = {Iris Agresti  and Davide Poderini  and Beatrice Polacchi  and Nikolai Miklin  and Mariami Gachechiladze  and Alessia Suprano  and Emanuele Polino  and Giorgio Milani  and Gonzalo Carvacho  and Rafael Chaves  and Fabio Sciarrino },
    title = {Experimental test of quantum causal influences},
    eprint = "2108.08926",
    archivePrefix = "arXiv",
    primaryClass = "quant-ph",
    month = "8",
    year = "2021",
    journal = {Science Advances},
    volume = {8},
    number = {8},
    pages = {eabm1515},
    year = {2022},
    doi = {10.1126/sciadv.abm1515},
    URL = {https://www.science.org/doi/abs/10.1126/sciadv.abm1515},
}

@manual{SagePolyhedra,
  title        = {SageMath: Combinatorial and Discrete Geometry},
  author       = {{The Sage Development Team}},
  year         = {2025},
  organization = {SageMath},
  note         = {Polyhedra and combinatorial geometry module},
  url          = {https://doc.sagemath.org/html/en/reference/discrete_geometry/index.html}
}

@article{rovelli2025back,
	author = {Carlo Rovelli},
	doi = {10.1007/s10838-024-09680-x},
	journal = {Journal for General Philosophy of Science / Zeitschrift f\"{u}r Allgemeine Wissenschaftstheorie},
	number = {1},
	pages = {97--115},
	publisher = {Springer Verlag},
	title = {Back to Reichenbach: Back to Reichenbach},
	volume = {56},
	year = {2024}
}

@book{price2007causation,
    author={Price, Huw and Corry, Richard},
    title = {Causation, Physics, and the Constitution of Reality: Russell’s Republic Revisited},
    publisher = {Oxford University Press},
    year = {2007},
    month = {02},
    isbn = {9780199278183},
    doi = {10.1093/oso/9780199278183.001.0001},
    url = {https://doi.org/10.1093/oso/9780199278183.001.0001},
}

@article{gachechiladze2020quantifying,
    author = "Gachechiladze, Mariami and Miklin, Nikolai and Chaves, Rafael",
    title = "{Quantifying causal influences in the presence of a quantum common cause}",
    eprint = "2007.01221",
    archivePrefix = "arXiv",
    primaryClass = "quant-ph",
    doi = "10.1103/PhysRevLett.125.230401",
    journal = "Phys. Rev. Lett.",
    volume = "125",
    number = "23",
    pages = "230401",
    year = "2020"
}

@article{mogstad2018using,
author = {Mogstad, Magne and Santos, Andres and Torgovitsky, Alexander},
title = {Using Instrumental Variables for Inference About Policy Relevant Treatment Parameters},
journal = {Econometrica},
volume = {86},
number = {5},
pages = {1589-1619},
doi = {https://doi.org/10.3982/ECTA15463},
url = {https://onlinelibrary.wiley.com/doi/abs/10.3982/ECTA15463},
year = {2018}
}

@article{cavalcanti2022post,
    author = "Cavalcanti, Paulo J. and Selby, John H. and Sikora, Jamie and Galley, Thomas D. and Sainz, Ana Bel{\'e}n",
    title = "{Post-quantum steering is a stronger-than-quantum resource for information processing}",
    eprint = "2102.06581",
    archivePrefix = "arXiv",
    primaryClass = "quant-ph",
    doi = "10.1038/s41534-022-00574-8",
    journal = "npj Quantum Inf.",
    volume = "8",
    number = "1",
    pages = "76",
    year = "2022"
}

@article{sainz2020bipartite,
    author = "Sainz, Ana Bel{\'e}n and Hoban, Matty J. and Skrzypczyk, Paul and Aolita, Leandro",
    title = "{Bipartite post-quantum steering in generalised scenarios}",
    eprint = "1907.03705",
    archivePrefix = "arXiv",
    primaryClass = "quant-ph",
    doi = "10.1103/PhysRevLett.125.050404",
    journal = "Phys. Rev. Lett.",
    volume = "125",
    pages = "050404",
    year = "2020"
}

@article{angrist1996identification,
author = {Joshua D. Angrist and Guido W. Imbens and Donald B. Rubin},
title = {Identification of Causal Effects Using Instrumental Variables},
journal = {Journal of the American Statistical Association},
volume = {91},
number = {434},
pages = {444--455},
year = {1996},
publisher = {Taylor \& Francis},
doi = {10.1080/01621459.1996.10476902},
URL ={https://www.tandfonline.com/doi/abs/10.1080/01621459.1996.10476902},
}

@article{villegas2024nonlocality,
  author={Villegas-Aguilar, Luis and Polino, Emanuele and Ghafari, Farzad and Quintino, Marco T{\'u}lio and Laverick, Kiarn T and Berkman, Ian R and Rogge, Sven and Shalm, Lynden K and Tischler, Nora and Cavalcanti, Eric G and Slussarenko, Sergei and Pryde, Geoff J.}, title = "{Nonlocality activation in a photonic quantum network}",
  eprint = "2309.06501",
  archivePrefix = "arXiv",
  primaryClass = "quant-ph",
  doi = "10.1038/s41467-024-47354-w",
  journal = "Nature Commun.",
  volume = "15",
  number = "1",
  pages = "3112",
  year = "2024"
}

@inproceedings{bonet2001instrumentality,
author = {Bonet, Blai},
title = {Instrumentality tests revisited},
year = {2001},
isbn = {1558608001},
publisher = {Morgan Kaufmann Publishers Inc.},
address = {San Francisco, CA, USA},
booktitle = {Proceedings of the Seventeenth Conference on Uncertainty in Artificial Intelligence},
pages = {48–55},
numpages = {8},
location = {Seattle, Washington},
series = {UAI'01},
eprint={1301.2258},
archivePrefix={arXiv},
primaryClass={cs.AI},
url={https://arxiv.org/abs/1301.2258},
}

@phdthesis{woodhead2014imperfections,
  title={Imperfections and self testing in prepare-and-measure quantum key distribution},
  author={Woodhead, Erik},
  year={2014},
  school={Ph. D. thesis, Universit{\'e} libre de Bruxelles}
}

@misc{polino2026network,
    author = "Polino, Emanuele and Poderini, Davide and Minati, Giorgio and Rodari, Giovanni and Chaves, Rafael and Sciarrino, Fabio",
    title = "{Network Nonlocality with Separable Measurements}",
    eprint = "2604.11910",
    archivePrefix = "arXiv",
    primaryClass = "quant-ph",
    month = "4",
    year = "2026"
}

@article{cabello2005bell,
    author = "Cabello, Adan and Feito, Alvaro and Lamas-Linares, Antia",
    title = "{Bell's inequalities with realistic noise for polarization-entangled photons}",
    eprint = "quant-ph/0509027",
    archivePrefix = "arXiv",
    doi = "10.1103/PhysRevA.72.052112",
    journal = "Phys. Rev. A",
    volume = "72",
    pages = "052112",
    year = "2005"
}

@article{canas2013detection,
    author = "Ca{\~n}as, Gustavo and Barra, Johanna F. and G{\'o}mez, Esteban S. and Lima, Gustavo and Sciarrino, Fabio and Cabello, Adan",
    title = "{Detection efficiency for loophole-free Bell tests with entangled states affected by colored noise}",
    eprint = "1206.2290",
    archivePrefix = "arXiv",
    primaryClass = "quant-ph",
    doi = "10.1103/PhysRevA.87.012113",
    journal = "Phys. Rev. A",
    volume = "87",
    pages = "012113",
    year = "2013"
}

@article{Hall2010,
    author = "Hall, Michael J. W.",
    title = "{Complementary contributions of indeterminism and signalling to quantum correlations}",
    eprint = "1006.3680",
    archivePrefix = "arXiv",
    primaryClass = "quant-ph",
    doi = "10.1103/PhysRevA.82.062117",
    journal = "Phys. Rev. A",
    volume = "82",
    pages = "062117",
    year = "2010"
}

@article{ChiribellasystemsCoexisting2024,
    author = "Chiribella, Giulio and Giannelli, Lorenzo and Scandolo, Carlo Maria",
    title = "{Bell Nonlocality in Classical Systems Coexisting with Other System Types}",
    eprint = "2301.10885",
    archivePrefix = "arXiv",
    primaryClass = "quant-ph",
    doi = "10.1103/PhysRevLett.132.190201",
    journal = "Phys. Rev. Lett.",
    volume = "132",
    number = "19",
    pages = "190201",
    year = "2024"
}
\clearpage
\newpage

\section*{Appendix/Supplementary material}

\appendix

\section{Operational Probabilistic Theories for generalized causal structures}
\label{app:OPTs}

To model nonclassical causal relations, we adopt the generalized Bayesian-network framework of Henson \textit{et al.}~\cite{henson2014theory}, formulated in the language of OPTs~\cite{chiribella2010probabilistic}. In an OPT, laboratory procedures are represented by tests that can be composed into circuits, with the theory assigning probabilities to the outcomes of any such composition. The corresponding notion of causality is weaker than the classical one: it only requires that the compositionality of the underlying OPT respects the causal structure in an appropriate way. This implies, e.g., that the statistics of a subset of variables does not depend on choices that are not among their causes, \textit{i.e.}, there is no-signaling from the causal future to the causal past. It also implies that correlations between variables with no directed causal path between them must be explained in terms of a common cause -- what in \cite{cavalcanti2014modifications} was called the Principle of Common Cause -- but does not imply \emph{Reichenbach's} Principle of Common Cause, which further requires a classical explanation whereby conditioning on the common cause eliminates the correlation. This framework naturally accommodates quantum theory, as well as more general probabilistic models.

A generalized causal structure is specified by a directed acyclic graph whose nodes are separated into observed and unobserved classes. The observed nodes encode classical random variables, and arrows coming out of observed nodes (observed edges) are associated with a classical copy of that node's variable. Unobserved nodes represent preparations or transformations of systems in the OPT: each arrow coming out of an unobserved node (unobserved edges) is associated with a distinct system $S_{X^i}$ in the theory, and the particular operation realized at an unobserved node may depend on its classical inputs. 

More formally, for each node $X^{(i)}$, and for each assignment $\mathrm{opa}\,x^{(i)}$ of its observed parents, one introduces a local \emph{test}
$\mathcal{T}_{x^{(i)}}(\mathrm{opa}\,x^{(i)})^{\mathrm{outU}X^{(i)}}_{\mathrm{incU}X^{(i)}}$,
which transforms the composite system $\mathrm{incU}X^{(i)}$ entering the node from its incoming unobserved edges (if any) into the composite system $\mathrm{outU}X^{(i)}$ leaving the node through outgoing unobserved edges (if any). The subscript $x^{(i)}$ represents the outcome of the test. When $X^{(i)}$ is an observed node, the outcomes of the associated test $\mathcal{T}_{x^{(i)}}(\mathrm{opa}\,x^{(i)})_{\mathrm{incU}X^{(i)}}$ (which has no outgoing systems) coincide with the possible values $x^{(i)}$ of the variable $X^{(i)}$; when $X^{(i)}$ is an unobserved node, the test $\mathcal{T}(\mathrm{opa}\,x^{(i)})^{\mathrm{outU}X^{(i)}}_{\mathrm{incU}X^{(i)}}$ has a single trivial outcome (which can be omitted from the subscript). The observed joint distribution is obtained by composing all such local tests in the pattern dictated by the graph. 

A probability distribution over the observed variables is \emph{compatible} with the causal structure if and only if there exists an assignment of tests to the nodes satisfying the rules above such that it satisfies the generalized Markov condition~\cite{henson2014theory}:
\begin{equation}
\label{eq:gen_markov} P\bigl(x^{(1)},x^{(2)},\dots,x^{(n)}\bigr)
=
\prod_{i=1}^{m}
\mathcal{T}_{x^{(i)}}(\mathrm{opa}\,x^{(i)})^{\mathrm{outU}X^{(i)}}_{\mathrm{incU}X^{(i)}} .
\end{equation}
In the special case when all variables in the structure are observed, all tests reduce to conditional probabilities $P(x^{(i)}|\mathrm{opa}\,x^{(i)})$. In this way, the classical Markov condition is recovered, while the generalized graphical language can be used to describe nonclassical causal models.

For more details of the OPT theory-independent causal framework, we refer to Ref.~\cite {henson2014theory}.

\begin{figure}
   \centering  \includegraphics[width=0.5\linewidth]{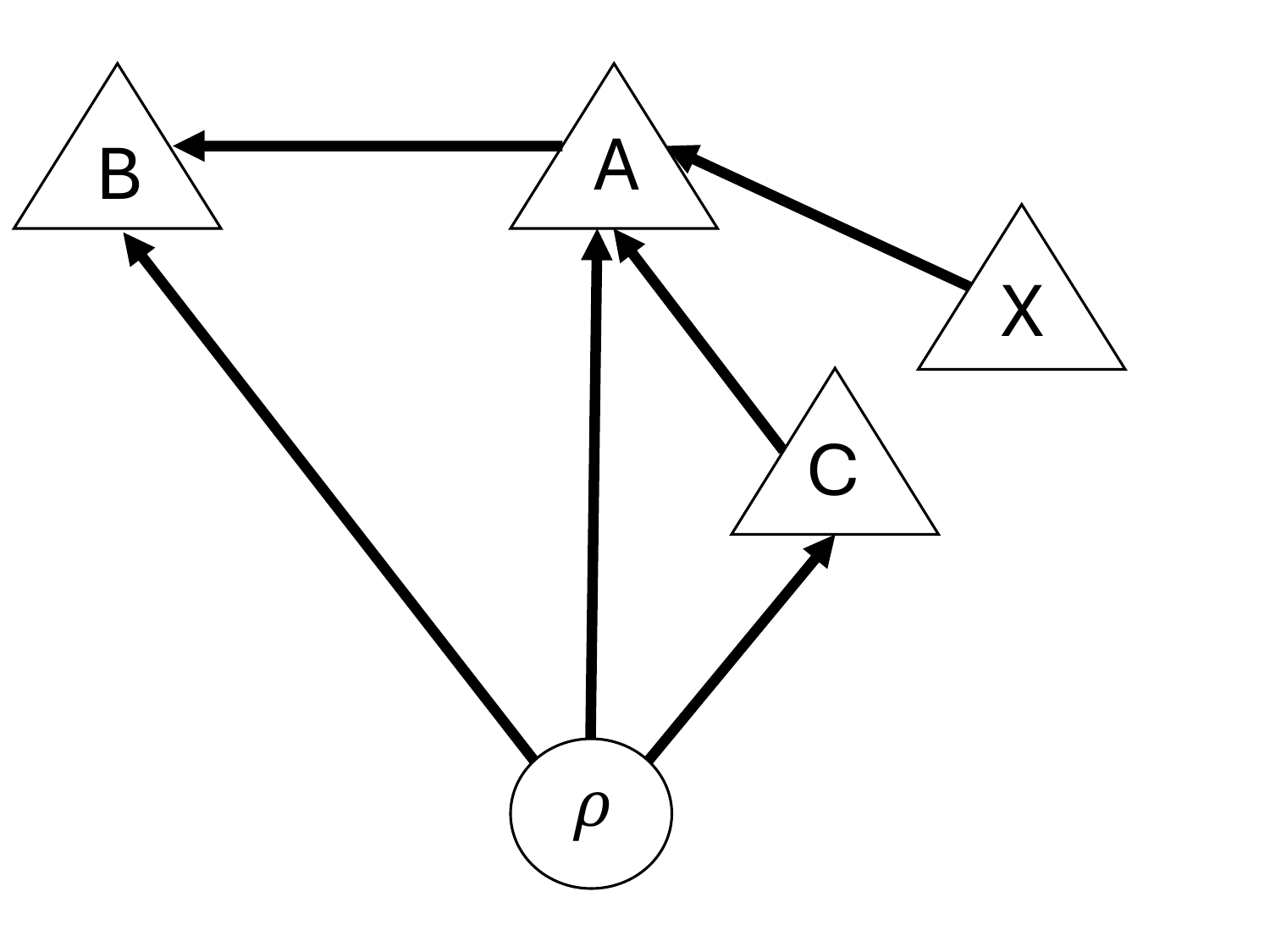}
   \caption{A generalized DAG, in the graphical notation of \cite{henson2014theory}, corresponding to the instrumental-based causal structure where the CF no-go result is exhibited.  The triangles represent observed nodes and have random variables assigned to them, while the circle represents a latent node which is associated with some general OPT systems but produces no outcomes. }
\label{fig:InstrumentalLFhlpGraph}
\end{figure}

\section{Necessary and sufficient constraints from the instrumental Wigner's friend causal structure\label{app:Necessary/SufficientConditionsFromHLP}}
In this section, we will show that Pearl's inequalities in Eq.~\eqref{eq:locality} and the freedom of choice constraints in Eq.~\eqref{eq:frewill} of the main text capture all the relevant causal constraints on $P(a,b,c|x)$ in the CF-instrumental scenario, and hence when paired with the tracking assumption in Eq.~\eqref{eq:aoe2}, they also capture all the relevant constraints for Causal Friendliness in the scenario, leading to the full characterization of the CF polytope presented in the main text.

In the case of the CF-instrumental scenario where the causal structure is as portrayed in Fig.~\ref{fig:InstrumentalLFhlpGraph} with the same notation used in~\cite{henson2014theory}, the generalized causal Markov condition of Eq.~\eqref{eq:gen_markov} gives 
\begin{align}
\nonumber
    P(a,b,c,x) &= \mathcal{T}_b(a)^{I}_{S_B}\mathcal{T}_a(x,c)^{I}_{S_A}\mathcal{T}_{c}()^I_{S_C}\mathcal{T}_x()^I_I \mathcal{T}_{\rho}()^{S_AS_BS_C}_{I} \\ 
    &=\mathcal{T}_b(a)^{I}_{S_B}\mathcal{T}_a(x,c)^{I}_{S_A}\mathcal{T}_{c}()^I_{S_C}\mathcal{T}_{\rho}()^{S_AS_BS_C}_{I} P(x),
\end{align}
where, on the second line, the fact that a test from a trivial system to another one is associated with a probability is used. Here $\mathcal{T}_{\rho}()^{S_AS_BS_C}_{I}$ is the preparation-test associated with the unobserved source, $\mathcal{T}_{c}()^I_{S_C}$ is the outcome-$c$ event at $C$, $\mathcal{T}_a(x,c)^I_{S_A}$ is the outcome-$a$ event at $A$, chosen as a function of the observed parents $x,c$, and $\mathcal{T}_b(a)^I_{S_B}$ is the outcome-$b$ event at $B$, chosen as a function of the observed parent $a$. The systems $S_A, S_B, S_C$ refer to the ones that connect the preparation test to the tests at the nodes $A, B, C$ respectively. 

 We can now compute the probability $P(c,x)$ of the free choice $x$ and the classical outcome $c$ by summing over the outcomes $a$ and $b$:
\begin{equation}
\begin{split}
& P(c,x) = \sum_{a,b} P(a,b,c,x) \\ &= P(x) \sum_{a,b} \mathcal{T}_b(a)^{I}_{S_B} \mathcal{T}_a(x,c)^{I}_{S_A} \mathcal{T}_{c}()^I_{S_C} \mathcal{T}_{\rho}()^{S_AS_BS_C}_{I}.     
\end{split} \end{equation} The sums over $a$ and $b$ are local coarse-grainings of complete observation-tests. Therefore, by OPT causality: 
\begin{equation} 
\sum_a \mathcal{T}_a(x,c)^I_{S_A} = \top_{S_A}^I, \qquad \sum_b \mathcal{T}_b(a)^I_{S_B} = \top_{S_B}^I\; , 
\end{equation} 
where, following the notation in~\cite{henson2014theory}, $\top_{S_A}^I$ and $\top_{S_B}^I$ are the unique deterministic effects on $S_A$ and $S_B$. 
It follows that 
\begin{equation} P(c,x) = P(x)\, \top_{S_B}^I \top_{S_A}^I \mathcal{T}_{c}()^I_{S_C} \mathcal{T}_{\rho}()^{S_AS_BS_C}_{I} = P(x)P(c), \end{equation} 
giving the No-Superdeterminism/Freedom of Choice condition of Eq.\eqref{eq:frewill}.  

Using $P(a,b,c,x) = P(a,b|c,x)P(c,x), $ the remaining relevant constraints are found from $P(a,b|x,c)$.
Note  first that (assuming $P(x,c)>0)$
\begin{equation}
\label{equation:conditionalonX,C}P(a,b\mid x,c) 
 = \frac{ \mathcal{T}_b(a)^{I}_{S_B} \mathcal{T}_a(x,c)^{I}_{S_A} \mathcal{T}_{c}()^I_{S_C} \mathcal{T}_{\rho}()^{S_AS_BS_C}_{I} }{ P(c) }. 
 \end{equation}

We will next show that any family of distributions $P(a,b|x,c)$  is compatible with the decomposition in Eq.~\eqref{equation:conditionalonX,C} if and only if they obey Pearl's inequalities as stated in Eq.~\eqref{eq:locality}. This is in fact, an immediate corollary of the results in Refs.~\cite{van2019quantum, henson2014theory} where constraints on the \emph{ordinary} instrumental scenario (i.e., the one which corresponds to the causal structure of Fig.~\ref{fig:InstrumentalLFhlpGraph} without the $C$ node) were investigated. The same techniques can be adapted to our case. 

\begin{figure}
    \centering
    \includegraphics[width=0.5\linewidth]{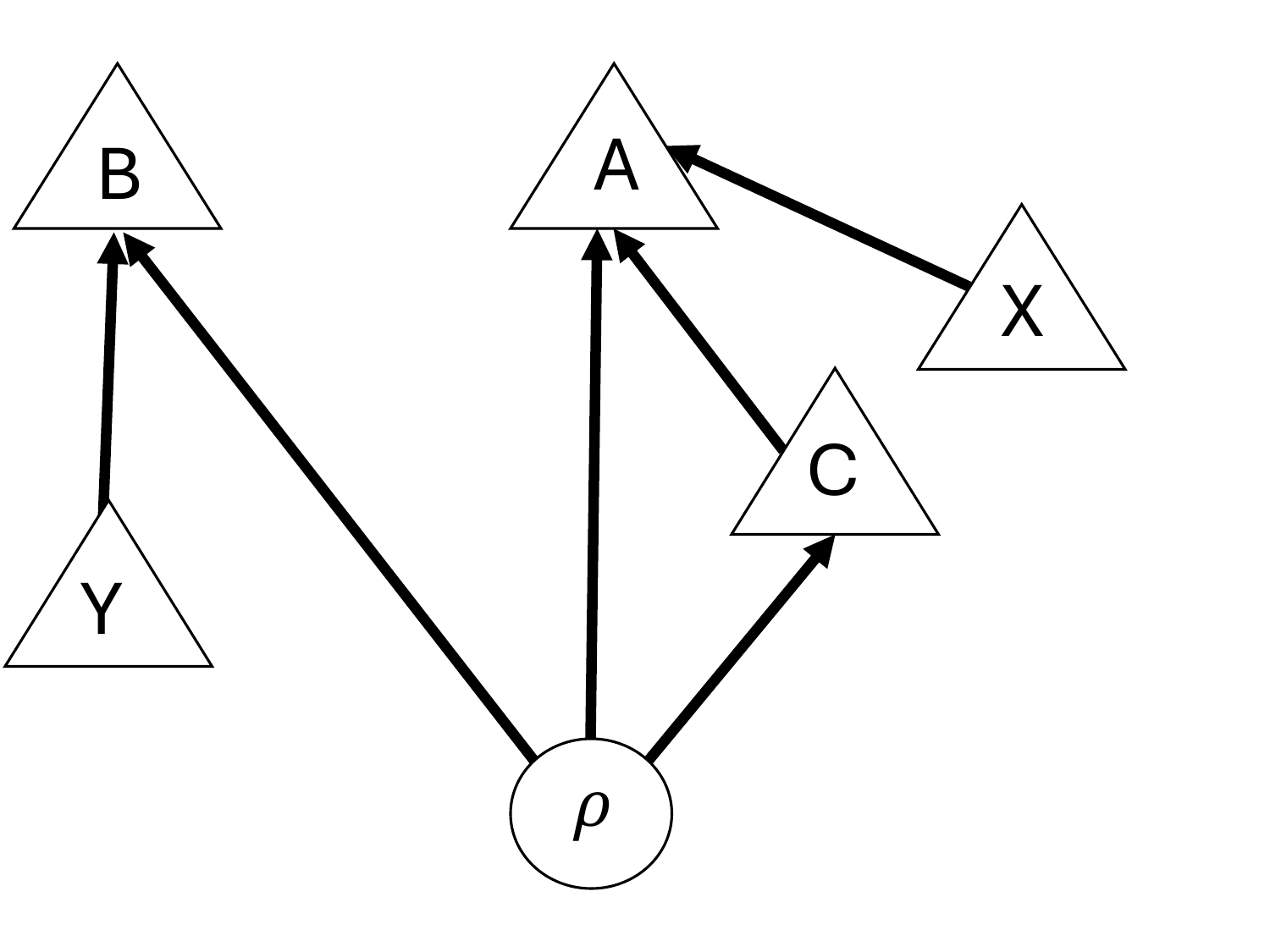}
    \caption{The generalized DAG corresponding to a canonical Local Friendliness scenario. In this picture, Alice and Bob perform test at spacelike separation, which corresponds to ruling out causal influence from the events on Bob's side to Alice's, and vice versa.  }
    \label{fig:CanonicalLFdag}
\end{figure}

The trick is to notice (following Refs.~\cite{henson2014theory, van2019quantum}) the close similarity of the constraints in Eq.~\eqref{equation:conditionalonX,C} to those arising in a \emph{canonical Local Friendliness scenario} (in the terminology of Ref.~\cite{haddara2025local}) where Alice and Bob are spacelike separated and the arrow from $A$ to $B$ is absent. The causal structure corresponding to this scenario is illustrated in Fig.~\ref{fig:CanonicalLFdag}. The generalized causal Markov condition of Eq.~\eqref{eq:gen_markov} implies that a distribution $Q(a,b,c,x,y)$ compatible with the causal structure in Fig.~\ref{fig:CanonicalLFdag} decomposes as
\begin{equation}\begin{split}
    Q(a,b,c,x,y) = \mathcal{T'}_b(y)&^{I}_{S_B'}\mathcal{T'}_a(x,c)^{I}_{S_A'}\mathcal{T'}_{c}()^I_{S_C'}\mathcal{T'}_{\rho}()^{S_A'S_B'S_C'}_{I} \\
    \times& Q(x)Q(y).
    \end{split}
\end{equation}

The analog argument shown for the case of the instrumental scenario implies that $Q(x,y,c)=Q(x)Q(y)Q(c)$ and so (with $Q(x,y,c)>0$) that 
\begin{align}
    Q(a,b|x,y,c) = \dfrac{\mathcal{T'}_b(y)^{I}_{S_B'}\mathcal{T'}_a(x,c)^{I}_{S_A'}\mathcal{T'}_{c}()^I_{S_C'}\mathcal{T'}_{\rho}()^{S_A'S_B'S_C'}_{I}}{Q(c)}.
\end{align}
The most general (i.e., complete) constraints on $Q(a,b|x,y,c)$ are well understood in these type of scenarios~\cite{bong2020strong, ying2024relating, haddara2025local, haddara2025properties}; they essentially demand that the $Q(a,b|x,y,c)$ satisfy parameter independence $\sum_a Q(a,b|x,y,c) = Q(b|y,c)$ and $\sum_b Q(a,b|x,y,c)=Q(a|x,c)$ for all $x,y,c$ which is to say that the collections of distributions $Q(a,b|x,y,c)$  are members of the no-signaling polytope for every $c$.

The observation that connects the above discussion back to the instrumental scenario is that since the tests and systems are allowed to be arbitrary, a comparison between the two scenarios \emph{within the same theory} allows one to equate the systems $S_A, S_B, S_C$ in the instrumental case with the primed systems in the canonical scenario, and permits the applications of the same tests in both cases, so that $\mathcal{T}'_c()_{S_C'}^I = \mathcal{T}_c()_{S_C}^I$ and so on. Therefore, without loss of generality, the distributions $Q(a,b|x,y,c)$ are compatible with generalized causal Markov in the scenario depicted in Fig.~\ref{fig:CanonicalLFdag} if and only if 
\begin{align}
    Q(a,b|x,y,c) = \dfrac{\mathcal{T'}_b(y)^{I}_{S_B} \mathcal{T}_a(x,c)^{I}_{S_A} \mathcal{T}_{c}()^I_{S_C} \mathcal{T}_{\rho}()^{S_AS_BS_C}_{I} }{ P(c)}. \label{Equation:ConditionalConstraintsOnNStests}
\end{align}
Now it is evident that if the cardinality of the set $Y$ is at least as great as that of $A$, it is possible to reproduce exactly the type of constraints on $P(a,b|x,c)$ as expressed in Eq.~\eqref{equation:conditionalonX,C} by considering the constraints on distributions $Q(a,b|x,y,c)$ as in Eq.~\eqref{Equation:ConditionalConstraintsOnNStests} and setting $\mathcal{T}'_b(y=y_a)=\mathcal{T}_b(a)$ for some $y_a\in Y$ (since the labels are arbitrary, without loss of generality $y_a=a$). The claim \emph{is not} that such a ``choice'' of $y$ can be realized in the canonical scenario given the output $a$ (since communication is excluded), but rather that the constraints in Eq.~\eqref{equation:conditionalonX,C} are inherently built in on the subsets of the distributions of the behaviours $Q(a,b|x,y,c)$, namely the ones of the form $Q(a,b|x,y=a,c)$, so that identification 
\begin{align}
      \begin{split}
      \label{eq:NS_To_Instu_condition}
  P(a,b|x,c)=&Q(a,b|x,y=a,c)\;\\
 &\forall \; x,\;a,\;b,\;  c\;
   \end{split} 
\end{align}
is possible. Eq.~\eqref{eq:NS_To_Instu_condition} essentially states that the valid instrumental distributions $P(a,b|x,c)$ are exactly those that can be recovered from behaviors $Q(a,b|x,y,c)$ by a mapping that only includes the probabilities where $y=a$. 
This map can also be directly seen by using the diagrammatic representations of the generalized causal structures~\cite{henson2014theory}: the diagram of the causal structure in Fig.~\ref{fig:InstrumentalLFhlpGraph} matches exactly with the one corresponding to the causal structure in Fig.~\ref{fig:CanonicalLFdag} when the condition $y=a$ is set.
In Ref.~\cite{van2019quantum}, it was pointed out that this map essentially amounts to projecting the no-signaling polytope (where the distributions $Q(a,b|x,y,c)$ live) onto a subspace, yielding another polytope. 

As we show in Appendix~\ref{app:pearlsunique}, the (non-trivial) facets of the projected polytope, which therefore capture the complete constraints for the instrumental distributions $P(a,b|x,c)$ are exactly the Pearl's inequalities (the case for the variables cardinalities $|X|=3, |A|=|B|=2$, was already shown in Ref.~\cite{van2019quantum}), which completes the proof.

As an immediate corollary, the CF-polytope is fully characterized by the causal constraints which are obeyed by all OPTs in the instrumental-based scenario, which are expressed in terms of Pearl's inequalities in Eq.~\eqref{eq:locality} and Freedom of Choice in Eq.~\eqref{eq:frewill}, and the tracking constraint in Eq.~\eqref{eq:aoe2} which is enforced by the specific reading of the measurement labelled by $X$. Let us also mention that the tracking constraint can also be directly baked into the distributions $Q(a,b|x,y,c)$ by demanding that they belong to the partially deterministic polytope~\cite{haddara2025local, haddara2025properties} corresponding to the canonical Local Friendliness scenario of Fig.~\ref{fig:CanonicalLFdag}, and projecting those to the instrumental scenario via Eq.~\eqref{eq:NS_To_Instu_condition}. By either approach, the complete characterization of the CF instrumental polytope presented in Sec.~\ref{Section:StrongQuantumNonclassicalityintheInstrumentalScenario} is obtained.

\section{Uniqueness of Pearl's inequalities for the OPT instrumental scenario}
\label{app:pearlsunique}
Pearl's inequalities (conditioned on $c$) in Eq.~\eqref{eq:locality} of the main text can be compactly expressed as 

\begin{align}
\label{Equation:PearlsInequalitiesInMaxForm} \max_a \sum_b \max_x P(a,b|x,c) \leq 1.
\end{align}
This form of the inequalities recovers the specific expression in Eq.~\eqref{eq:locality} of the main text with the fixed cardinalities, but it can also be evaluated in more general settings as we do in this section. 

In appendix \ref{app:Necessary/SufficientConditionsFromHLP} it was established that the distributions $P(a,b,c,x)$ defined over the instrumental-based scenario are those which factor as
\begin{align}
    P(a,b,c,x) = P(a,b|x,c)P(x)P(c)
\end{align}
and where the conditional distributions $P(a,b|x,c)$ are those that can be obtained from a no-signaling behaviour $Q(a,b|x,y,c)$ via the map in Eq.~\eqref{eq:NS_To_Instu_condition}  as $P(a,b|x,c)= Q(a,b|x, y=a,c)$. Here the set $Y$ of Bob's inputs is without loss of generality equal to Alice's outcome set $A$. In this section, we fully characterize the set of OPT instrumental distributions $P(a,b|x,c)$ compatible with such constraints, demonstrating the following theorem:

\begin{theorem}
The OPT instrumental distributions $P(a,b|x,c)$, for any finite cardinality of $a,b, x,c$, are exactly those satisfying positivity, normalization, and Pearl's instrumental inequalities in the form of Eq.~\eqref{Equation:PearlsInequalitiesInMaxForm}.
\end{theorem}
It is sufficient to show that the relation 
\begin{align}
    P(a,b|x,c) = Q(a,b|x,y=a,c)\label{Equation:RelationOfInstrumentalAndNSInProof}
\end{align}
implies that Pearl's inequalities in the form \eqref{Equation:PearlsInequalitiesInMaxForm} are satisfied by the instrumental distributions $P(a,b|x,c)$ and conversely, that if Pearl's inequalities hold for $P(a,b|x,c)$ then there exists a no-signaling distribution $Q(a,b|x,y,c)$ such that the relation \eqref{Equation:RelationOfInstrumentalAndNSInProof} holds. In the forthcoming discussion, we refer to Eq.~\eqref{Equation:RelationOfInstrumentalAndNSInProof} as ``the instrumental constraints''. 

\paragraph*{$[ \Longrightarrow]$ Instrumental constraints imply Pearl's inequalities.}

Let $Q(a,b|x,y,c)$ be a no-signaling behaviour which recovers the instrumental distributions $P(a,b|x,c)$ via Eq.~\eqref{Equation:RelationOfInstrumentalAndNSInProof}. Fix a value $y=\alpha$ of Bob's input. Since $Q$ is no-signaling from Alice to Bob, the marginal distribution of Bob's outcome
for the input $y=\alpha$ cannot depend on $x$. Hence, for each
$b$, there exists a number $m_{\alpha b}^c$, independent of $x$, such that
\begin{equation}
    \sum_a Q(a,b|x,y=\alpha,c) = m_{\alpha b}^c
    \qquad
    \forall \,x .     
\end{equation}
In particular, among the terms in the sum over Alice's output $a$, there is
the term with $a=\alpha$. Using the defining relation between
$P$ and $Q$, we have
 
\begin{equation}
\begin{split}
    m_{\alpha b}^c
    =&
    \sum_a Q(a,b|x,y=\alpha,c)
    \geq
    Q(\alpha,b|x,y=\alpha,c) \\
    =&
    P(\alpha,b|x,c) 
    \hspace{0.3cm}
    \forall\, x,c \;.    
\end{split}
\end{equation}
Thus in particular
\begin{equation}
    m_{\alpha b}^c \geq \max_x P(\alpha,b|x,c).
\end{equation}

Moreover, for each fixed $\alpha,c$, the numbers $m_{\alpha b}^c$ by definition give a normalized
probability distribution over $b$:
$ \sum_b m_{\alpha b}^c=\sum_b \sum_a Q(a,b|x,\alpha,c) =1$.
Combining the previous two observations gives
\begin{equation}
    \sum_b \max_x P(\alpha,b|x,c)
    \leq
    \sum_b m^c_{\alpha b}
    =
    1.  
\end{equation}
Since this conclusion holds for every value $\alpha$, it holds in particular for the $\alpha$ which maximizes the sum, so that one obtains Pearl's inequalities:
\begin{equation}
   \max_{a} \sum_b \max_x P(a,b|x,c) \leq 1 \;.
\end{equation}
This direction was shown by Henson et al.~\cite{henson2014theory}.\\

$[\Longleftarrow]$ \textit{ Pearl's inequalities imply the instrumental constraints.}\\
Conversely, suppose that $P(a,b|x,c)$ satisfies Pearl's inequalities so that
\begin{equation}
\sum_b \max_x P(\alpha,b|x,c)\le 1
\qquad \forall \alpha .
\end{equation}
For each value $\alpha$ of Alice's output $a$, we define a probability distribution
$\{m_{\alpha b}^c\}_b$ such that
\begin{equation}
m_{\alpha b}^c\ge \max_x P(\alpha,b|x,c),
\qquad 
\sum_b m_{\alpha b}^c=1 .
\end{equation}
Such $m_{\alpha b}^c$ exist precisely because Pearl's inequality holds. As a concrete example, if $\sum_b \max_x P(\alpha,b|x,c) =1$, the identification $m^c_{\alpha b}=\max_x P(\alpha,b|x,c)$ is sufficient, otherwise if   $\sum_b \max_x P(\alpha,b|x,c) =k <1$, one can define  $m_{\alpha b'}^c =\max_x P(\alpha,b'|x,c) +1-k$ for some $b'$, and $m^c_{\alpha b }=\max_x P(\alpha,b|x,c)$ for $b\neq b'$, say.

We now want to construct a bipartite no-signaling distribution $Q(a,b|x,y,c)$ satisfying
$Q(a,b|x,y=a,c)=P(a,b|x,c)$. \\
First, with $y=\alpha$, set:
\begin{equation}
Q(\alpha,b|x,y=\alpha,c)=P(\alpha,b|x,c).
\end{equation}

We will show that the entries
$Q(a,b|x,y=\alpha,c)$ with $a\neq \alpha $ can always be chosen
consistently. For fixed $x$ and $y=\alpha$, regard these
entries as elements $M^{x\alpha c}_{ab}$ of an unknown non-negative matrix $M^{x\alpha c}$ whose rows are
labeled by $a\neq\alpha$ and whose columns are labeled by
$b$. We require its row sums to be
\begin{equation}
R^{x\alpha c}_a=\sum_bQ(a,b|x,y=\alpha,c):=P(a|x,c),
\qquad a\neq\alpha,
\end{equation}
and its column sums to be
\begin{equation}
C^{x\alpha c}_b=\sum_{a\neq\alpha}Q(a,b|x,y=\alpha,c):=m^c_{\alpha b}-P(\alpha,b|x,c)\;.
\end{equation}
The row sums are non-negative because they are probabilities.
The column sums are also non-negative because
\begin{equation}
m^c_{\alpha b}\ge \max_x P(\alpha,b|x,c)\ge P(\alpha,b|x,c)\; \forall \alpha,b, x,c.
\end{equation}
For these requirements to be consistent, it is necessary that the sum over all the rows equals the sum over all the columns. This is easily verified as the total row sum gives:
\begin{equation}
\sum_{a\neq\alpha}R^{x\alpha c}_a
=
\sum_{a\neq\alpha}P(a|x,c)
=
1-P(\alpha|x,c)\;,
\label{eqapp:row}
\end{equation}
while
\begin{equation}
\begin{split}
&\sum_b C^{x\alpha c}_b
=
\sum_b\left[m^c_{\alpha b}-P(\alpha,b|x,c)\right]
= \\
&= 1-\sum_b P(\alpha,b|x,c)
=
1-P(\alpha|x,c)\;.
\end{split}
\label{eqapp:column}
\end{equation}
This consistency condition is also sufficient for the existence of a completion of the non-negative matrix $M^{x\alpha c}$ of the $Q(a,b|x,y=\alpha,c), a\neq \alpha$ with entries that are compatible with the constraints on the row and column sums in Eqs.~\eqref{eqapp:row}--\eqref{eqapp:column}. Indeed, as a concrete example, one can set 
\begin{equation} M_{ab}^{x\alpha c} := \begin{cases} \displaystyle \frac{ R_a^{x \alpha c} \; C_b^{x\alpha c} }{ 1-P(\alpha\mid x,c) }, & P(\alpha\mid x,c)<1, \\[2ex] 0, & P(\alpha\mid x,c)=1, \end{cases} \quad a\neq\alpha \,, 
\end{equation} 
which is always possible. Finally, we observe that every matrix $M^{x\alpha c}$ compatible with the row and column constraints gives a valid extension to a no-signaling distribution $Q(a,b|x,y,c)$. Namely since now by construction
\begin{align} \nonumber
\sum_a& Q(a,b|x,y=\alpha,c)= C^{x\alpha c}_b +Q(\alpha, b|x,y=\alpha ,c) \\
&=[m_{\alpha b}^c - P(\alpha, b|x,c) ]+P(\alpha, b|x, c) =m^c_{\alpha b}\;,
\end{align}
which is independent of $x$, so No Signaling (NS) from Alice to Bob holds. For $a=\alpha$ we have
\begin{equation}
\sum_b Q(\alpha,b|x,y=\alpha,c)
=
\sum_b P(\alpha,b|x,c)
=
P(\alpha|x,c)\;,
\end{equation}
while for $a\neq\alpha$, the prescribed row sums give
\begin{equation}
\sum_b Q(a,b|x,y=\alpha,c)=P(a|x,c)\;.
\end{equation}
Thus, for any value of $a$, also no-signalling from Bob to Alice holds. Therefore $Q$
is a bipartite NS extension of $P$, and by construction
\begin{equation}
Q(a,b|x,y=a,c)=P(a,b|x,c),
\end{equation}
establishing the claim.

Note that we include $c$ among the conditioning variables in the probabilities to account for the presence of the variable $C$ in the instrumental Wigner's friend causal structure shown in Fig.~\ref{fig:InstrumentalLFhlpGraph}. However, exactly the same construction also applies to the standard instrumental scenario, where the node $C$ is absent. Therefore, Pearl's inequalities in the form of Eq.~\eqref{Equation:PearlsInequalitiesInMaxForm} (without the conditioning on $c$) give the necessary and sufficient constraints for distributions $P(a,b|x)$ to be compatible with the generalized causal Markov condition in the ordinary instrumental scenario.

\section{Exclusivity principle as an alternative nonclassical principle for generalized causality}
\label{app:ExclusivityrelationOPT}

In this appendix, we focus on the relation between CF and the exclusivity principle in the instrumental scenario.

The exclusivity graph approach~\cite{cabello2014graph,acin2015combinatorial}, suitably extended, is an alternative technique for characterizing correlations arising from sets of hybrid assumptions in arbitrary scenarios~\cite{rodari2023characterizing}. 
This approach is particularly convenient when one wants to study or find inequalities satisfied by the models under consideration, without fully characterizing the polytope.

For the instrumental scenario considered in this work, we have dichotomic outcomes for both $A$ and $B$, i.e., $a,b\in\{-1,+1\}$, and the measurement choice value  $x\in\{0,1,2\}$. We apply the CF constraints, where the outcome $a$, conditioned on the value $c$, is deterministic (more generally, classical) only for the measurement choice $X=0$. 
Here, each possible observable event $e\,=\,(a,b|x)$ represents a node of the exclusivity graph, and its exclusivity relation with another event $e'\,=\,(a',b'|x')$ is represented by an edge and defined as: 

\begin{equation*}
(a,b | x) \sim (a',b' | x')
\;\Longleftrightarrow\;
\left\{
\begin{array}{l}
a \neq a',\; x = x',\\[2pt]
\text{or}\\[2pt]
b \neq b',\; a = a'\; .
\end{array}
\right.  
\end{equation*}

\begin{figure*}
   \centering  \includegraphics[width=0.8\linewidth]{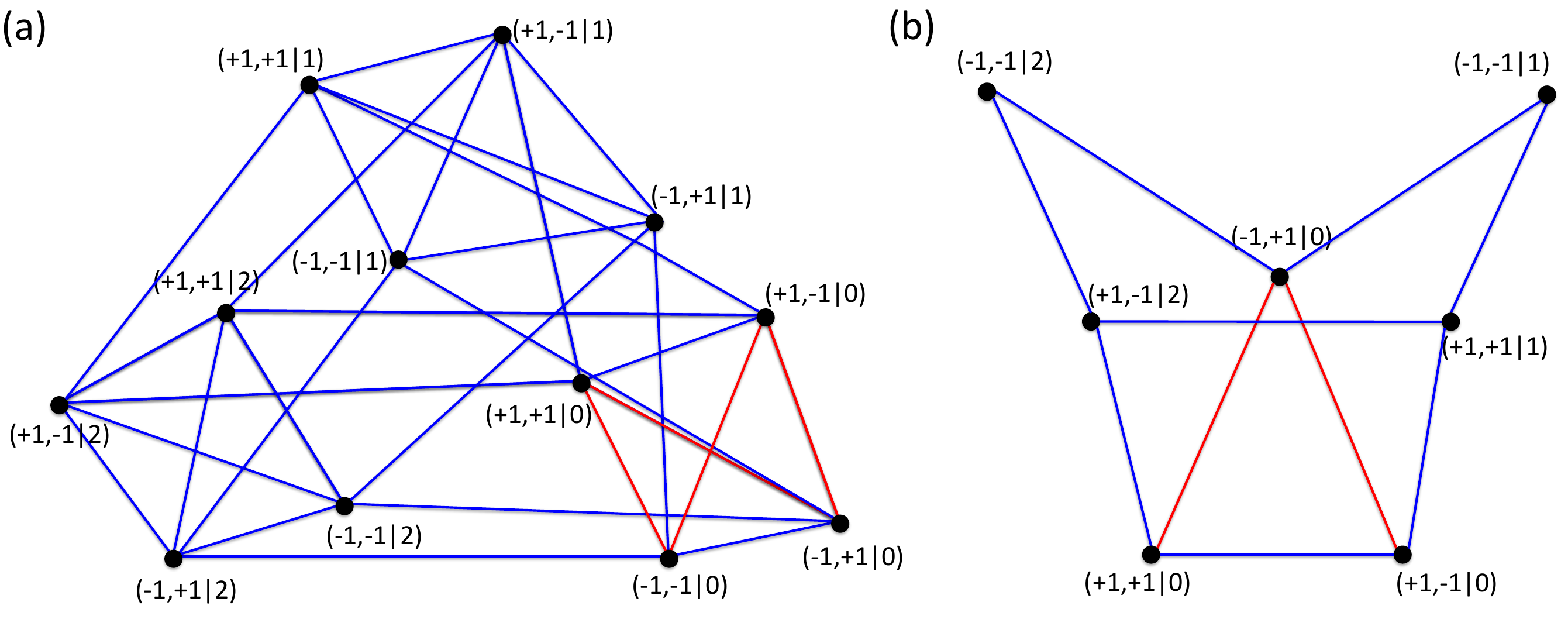}
   \caption{\textbf{(a)} Exclusivity graph of the instrumental Wigner's friend scenario where the red edges are those connecting events involved in the AOE assumption, while the blue ones are standard exclusivity edges. \textbf{(b)} Induced subgraph involving the events involved in the genuine facet inequality in Eq.~\eqref{eq:genuinelf}.}
\label{fig:graphs}
\end{figure*}

Notably, the exclusivity principle applied to the instrumental exclusivity graph yields exactly Pearl's inequalities in Eq.~\eqref{eq:locality}, thereby fully characterizing the corresponding OPT relations.
The complete exclusivity graph for the considered instrumental Wigner's friend scenario is depicted in Fig.~\ref{fig:graphs}a. 
Here, in order to impose different constraints to the events (the AOE and tracking assumption only constrains the events with the choice $X=0$) there are two kinds of edges: those relative to the constraint of generalized causality in Eq.~\eqref{eq:locality} (blue edges in Fig.~\ref{fig:graphs}) which are only constrained by the exclusivity principle, and those connecting exclusive events involved in the AOE relation in Eq.~\eqref{eq:aoe}, which have deterministic (more generally, classical) outcomes occurring when the choice $X=0$ (red edges in Fig.~\ref{fig:graphs}). The approach in Ref.~\cite{rodari2023characterizing} can handle this kind of graph with hybrid assumptions, retrieving constraints on the observed probabilities from the CF assumptions.  
From the complete graph, we can select a subgraph (the exclusivity graph of the subset of nodes and their relative exclusivity relations) shown in Fig.~\ref{fig:graphs}(b), which represents the graph of the events involved in the genuine facet in Eq.~\eqref{eq:genuinelf}. By suitably calculating the graph invariants, we can directly compute the bounds of linear inequalities for the CCM, CF, and OPT models in the instrumental Wigner's friend scenario, without explicitly characterizing the full polytopes.

Given a graph $G$ and weights $w$ on the events and a linear combination of the probabilities $I_{inst}(G,w)$, one can always compute:
$\alpha(G,w)$, known as the independence number of a graph, provides a tight bound for classical causal models (CCMs) respecting classical causality; 
$\theta(G,w)$, called Lova\`sz Theta of a graph, provides an upper quantum bound (Q) for the inequality ~\cite{knuth1993sandwich,grotschel1986relaxations} when it is mapped onto an exclusivity graph $G$ and a set of weights $w$ ~\cite{cabello2013simple, cabello2013basic, poderini2020exclusivity}; 
$\alpha_{EP}(G,w)$ is the bound for OPTs obeying the exclusivity principle, which in standard bipartite scenarios corresponds to NS bounds~\cite{acin2015combinatorial,cabello2014graph};
$\alpha_{CF}(G,w)$ accounts for the two different types of edges from CF assumptions applied to the instrumental Wigner's friend graph and can be calculated using the approach in Ref.~\cite{rodari2023characterizing}.

In general, the relation holds  $I(G,w) \overset{\text{CCM}}{\leq} \alpha(G,w)  \overset{\text{Q}}{\leq} \theta(G,w)\overset{\text{EP}}{\leq} \alpha_{EP}(G,w)$
The following inequalities also hold 
 $I(G,w) \overset{\text{CCM}}{\leq} \alpha(G,w)  \overset{\text{CF}}{\leq} \alpha_{CF}(G,w)\overset{\text{EP}}{\leq} \alpha_{EP}(G,w)$.

Nevertheless, the relation between $\alpha_{CF}$ and $\theta(G,w)$ is not fixed, but depends on the particular inequality considered, given that there is no inclusion relation between CF and quantum correlation sets.

\section{Correlation sets and axes expressions}
\label{app:plotexpressions}
The expressions on the axes of the 2-D polytopes' projection in Fig.~\ref{fig:results} are the following: 
\begin{equation}
    \begin{split}
        &\mathcal{E}_1 = 2.7 \cdot (\mathcal{E}'_1 + \mathcal{E}'_2)\\
        &\mathcal{E}_2 = 2 \cdot (- \mathcal{E}'_1 + \mathcal{E}'_2)
    \end{split}
\end{equation}
with:

\begin{equation}
    \begin{split}
        \mathcal{E}'_1 = &-\powerset(-1,+1|0) + \powerset(+1,-1|1) + \powerset(-1,+1|1)\\
        &+ \powerset(+1,+1|2) + \powerset(-1,+1|2)
    \end{split}
\end{equation}
and
\begin{equation}
    \begin{split}
        \mathcal{E}'_2 = & \;0.2\cdot\powerset(-1,-1|0) + \powerset(+1,+1|0) + \powerset(-1,-1|2)\\
        &+ \powerset(+1,-1|2) + 2\cdot\powerset(-1,+1|0) + \powerset(+1,+1|1)\\
        &+ \powerset(-1,-1|1) + \powerset(+1,-1|0)
    \end{split}
\end{equation}

The choice of the expressions themselves was optimized to showcase the location of the two experimental points; one ($X_A$) lying entirely within the CF polytope but outside the CCM one, while the other ($X_B$) lying outside both the CF and the CCM polytopes, as is evident in Fig.~\ref{fig:results}.
The polytopes are computed using the standard Python software SageMath with the Polyhedra library~\cite{SagePolyhedra}.

\section{Quantum strategy violating CF constraints and noise robustness}
\label{app:noiserobustness}

As mentioned previously in the main text, an optimal quantum strategy to find a violation of Eq.~\eqref{eq:genuinelf} requires us to consider the singlet state $\ket{\Psi} = (\ket{01} -\ket{10})/\sqrt{2}$ and the observables  $\hat A_0=-\sigma_z,\;
\hat A_1=(\sigma_z-\sigma_x)/\sqrt2,\;
\hat A_2=(\sigma_z+\sigma_x)/\sqrt2$ for Alice, and $\hat B_{+1}=\sigma_x,\;
\hat B_{-1}=-\sigma_z$ for Bob.

\begin{figure}
    \centering
    \includegraphics[width=\columnwidth]{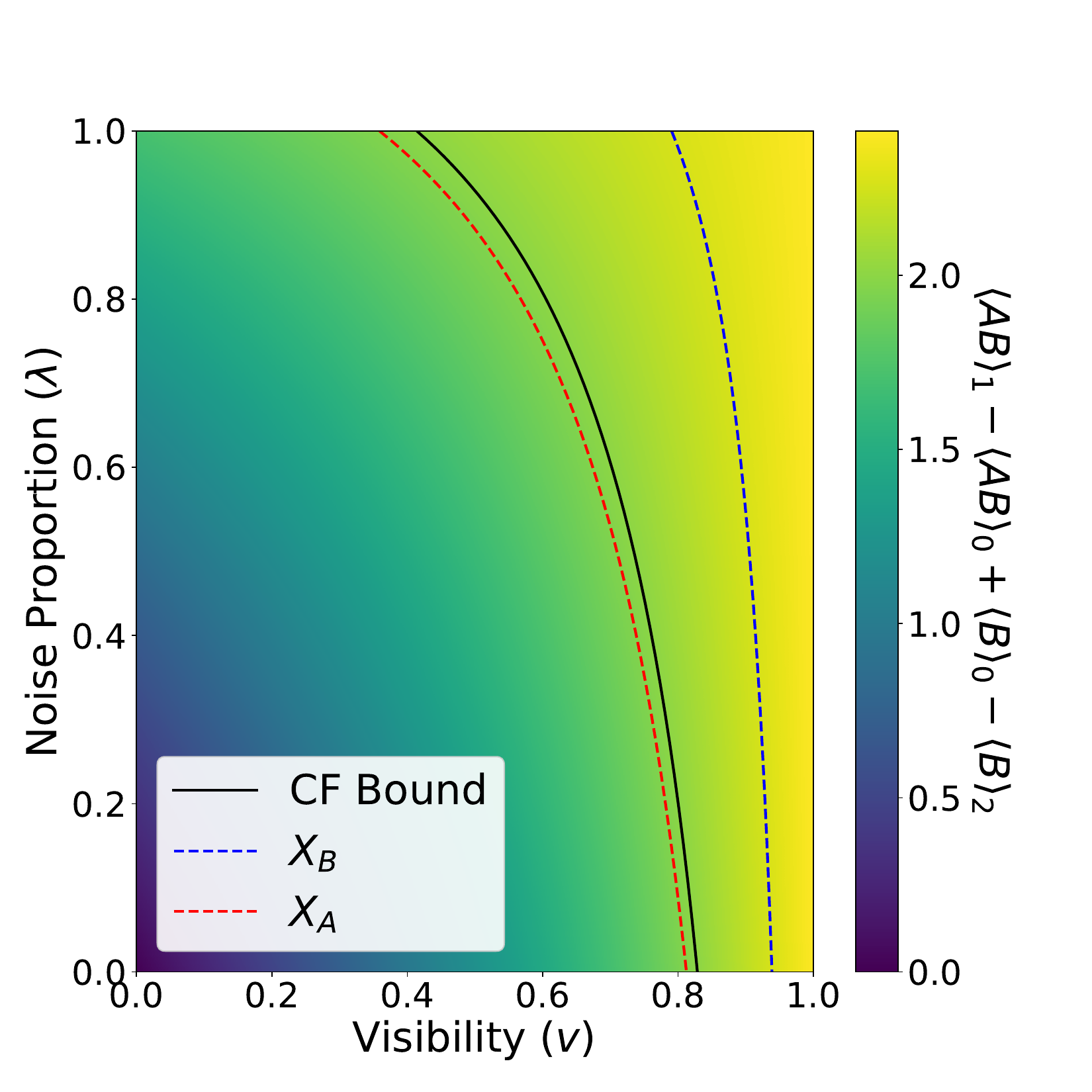}
    \caption{\textbf{Noise robustness of the genuine CF inequality.} The L.H.S of Eq.~\eqref{eq:genuinelf} has been calculated at 250,000 points and plotted as a heatmap by using the noisy model in Eq.~\eqref{eq:noisymodelvisibility} and the optimal measurement settings. The axes span values of the visibility $(v)$ and the noise proportion $(\lambda)$. The black contour marks the boundary between points that violate and satisfy the genuine CF inequality; the points to the right of the contour populate the region that violates Eq.~\eqref{eq:genuinelf}\ and is robust to noise. We also show the contours corresponding to the values of the L.H.S. of Eq.~\eqref{eq:genuinelf} from the experimental data (refer to Fig.~\ref{fig:results}): in blue is the point $X_B$ that corresponds to the experimental violation of the genuine CF facet, and in red is the point $X_A$, which has a value that lies outside of the CCM polytope but inside the CF one. It is possible to move from one regime to the other, for instance, by changing the parameters of the noise in the state in Eq.~\eqref{eq:noisymodelvisibility}: this is the strategy we implemented experimentally by enlarging the coincidence window to measure the point $X_A$.
    }
    \label{fig:noiserobustness}
\end{figure}
To demonstrate this experimentally, polarization-entangled photon pairs from a Spontaneous Parametric Down-Conversion (SPDC) source are used (refer to Sec.~\ref{sec:experiment}). Such sources are affected by two distinct kinds of noise~\cite{cabello2005bell, canas2013detection}, which leads to the adoption of the following noise model:
\begin{equation}\label{eq:noisymodelvisibility}
     \begin{split}
         \rho =  &v\cdot|\Psi\rangle\langle\Psi|+\\
         &+ (1 - v)\left[\lambda\frac{|HV\rangle\langle HV| + |VH\rangle\langle VH|}{2} + (1 - \lambda)\frac{\mathbb{I}}{4}\right]\;,
     \end{split}
\end{equation}
Here, $|\Psi\rangle = (|HV\rangle - |VH\rangle)/\sqrt{2}$ is the pure singlet state that is ideally being produced by the source. The \textit{white noise} is responsible for the appearance of the maximally mixed state, which corresponds to an overall depolarization of the state. The \textit{coloured noise} is associated with depolarization in a specific direction, and due to reasons related to the generation of the $|HV\rangle$ and $|VH\rangle$ states by SPDC, this noise presents itself as a decoherence term, affecting the specific off-diagonal components of the density matrix. The \textit{visibility} $0 \le v \le 1$ parametrizes the portion of the quantum state that is ideal and noiseless, and the \textit{noise proportion} given by $0 \le \lambda \le 1$ is the colored noise fraction of the non-singlet mixture.

The parameters $v$ and $\lambda$ enable us to calibrate the constraints of this quantum strategy, and tuning their values also leads to an identification of the regime in which it is robust to this category of noise. Having fixed the singlet state and measurement operators to the ones mentioned before, we now consider the behavior of Eq.~\eqref{eq:genuinelf} as a function of these noise parameters (refer to Fig.~\ref{fig:noiserobustness}). Note the (black) contour that marks the boundary between violation and satisfaction of the inequality; it demarcates the regions and provides the limits of our quantum strategy: the lower bound on visibility goes from about $82.84\%$ when $\lambda = 0$, to about $41.42\%$ when $\lambda = 1$. The dashed contours in red and blue represent results obtained from experimental data.

It is important to highlight that this analysis is only gauging the robustness of the noise created by the SPDC source and quantifying how far the experimentally generated state can  be from the ideal scenario and still constitute a violation of the instrumental inequality.  More precisely, this calibrates the limit on the error of the state generation, and there are other sources of error subsequently that will lead to a less-than-optimal result; for instance, the switching error of the optical switch (see discussion in Appendix~\ref{app:expdetails}).

\section{Experimental details and outcome assignment strategy}
\label{app:expdetails}

\begin{figure*}[ht!] \centering \subfloat[]{ \includegraphics[width=0.5\textwidth]{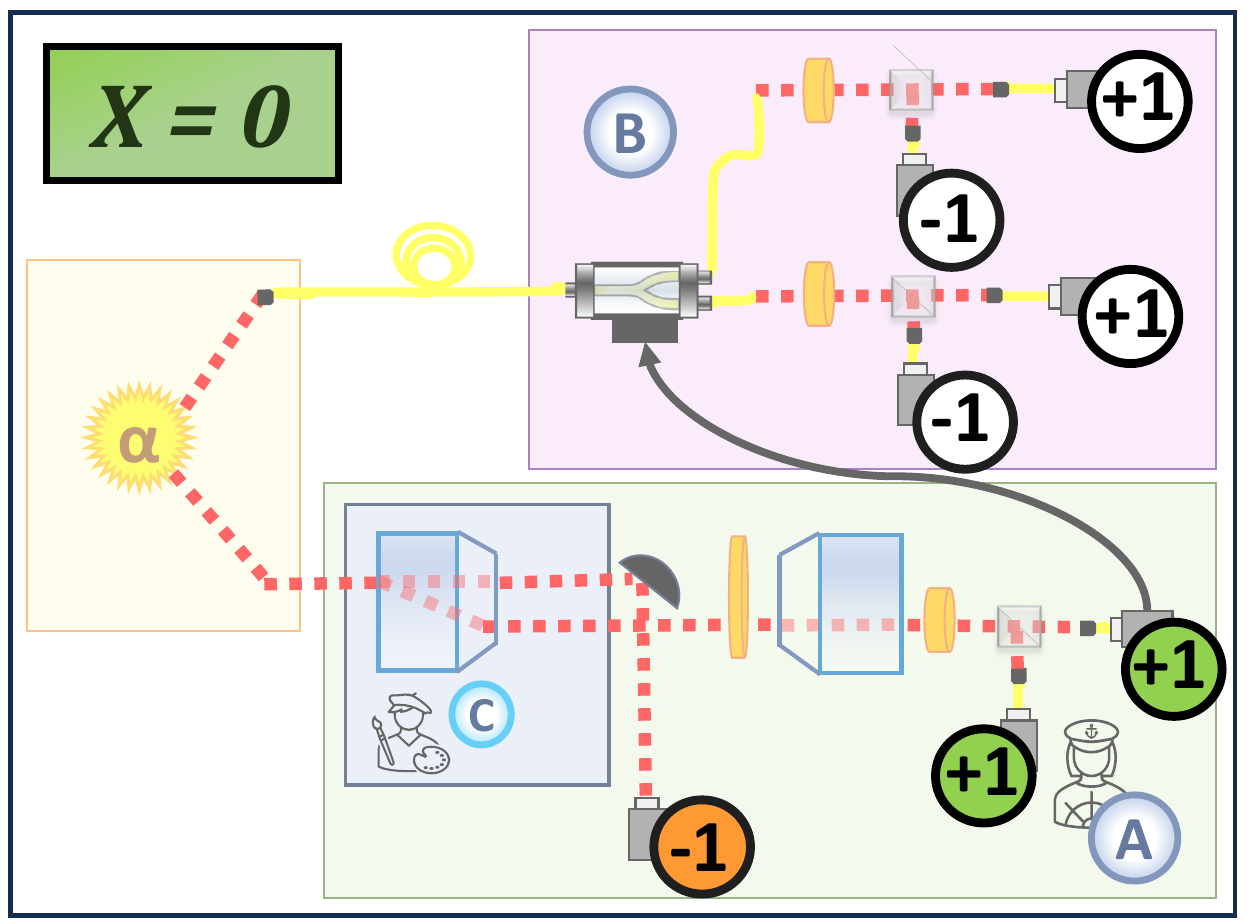} \label{fig:expsetuplabels0} }
\subfloat[]{ \includegraphics[width=0.5\textwidth]{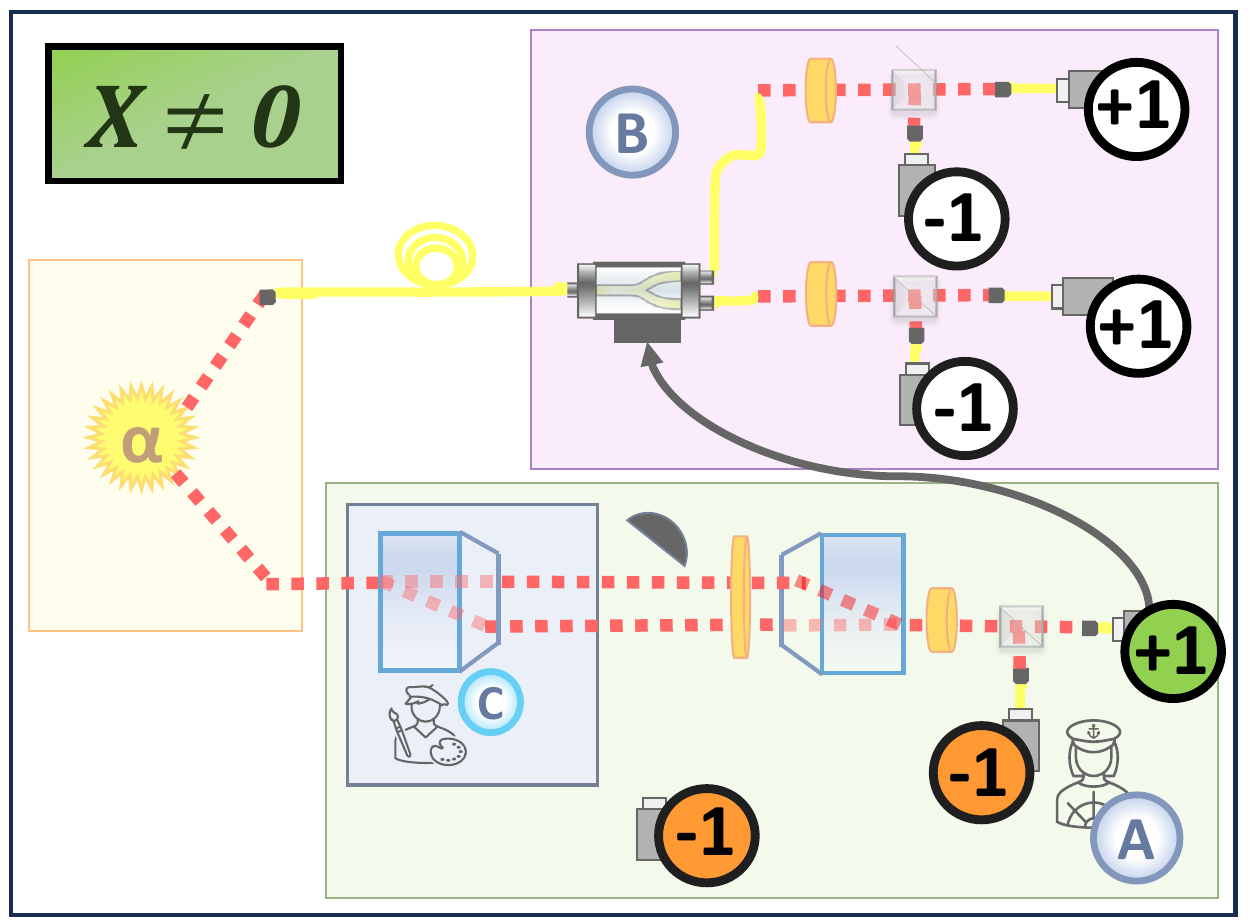} \label{fig:expsetuplabels12} } \caption{\textbf{Assignment of outcomes in the experimental setup.} Bob's measurement station performs fixed polarization measurements based on Alice's outcome, and the outcome labels are therefore fixed. By contrast, the outcome labels in Alice's station depend on the measurement choice $x$, which determines whether Alice measures the photon's path (corresponding to asking the friend for their outcome) or its polarization (corresponding to reversing the interaction and then performing a measurement). The corresponding labeling convention is illustrated in panel~\textbf{(a)} for $x=0$ and in panel~\textbf{(b)} for $x\neq0$.} 
\label{fig:expsetuplabels} \end{figure*}

The instrumental scenario is implemented in the photonic platform already described in  Fig.~\ref{fig:expsetup}\subref{fig:setup}; here, we specify the assignment of the measurement outcomes in Fig.~\ref{fig:expsetuplabels}, for both $x = 0$ and $x \neq 0$. 

Charlie always performs a $\sigma_z$ measurement, implemented by a beam displacer that couples polarization and path. When Alice decides to ask her friend the result of his measurement ($x = 0$), a motorized mirror is inserted, and the polarization of each photon is measured through its path, as shown in Fig.~\ref{fig:expsetuplabels}\subref{fig:expsetuplabels0}. Therefore, we assign the outcome $-1$ to the detector placed in front of the motorized mirror, which collects photons from the path corresponding to vertical polarization; consistently, we assign the outcome $+1$ to both detectors after Alice's PBS, which together collect photons coming from the other path. 
On the other hand, when $x \neq 0$, the friend's interaction with the system is reversed, and Alice's measurement is implemented in the polarization degree of freedom by an HWP and a PBS. The path of the photons for these measurement choices is shown in Fig.~\ref{fig:expsetuplabels}\subref{fig:expsetuplabels12}. In accordance with the previous case, the outcome $+1$ is again assigned to the detector collecting the photons transmitted by the PBS, which provides the feed-forward for Bob's measurement settings; Alice's remaining detectors are then assigned the outcome $-1$.
In contrast, Bob's measurement station consists of two static polarization measurements; the outcomes' labels are shown in both Fig.~\ref{fig:expsetuplabels}\subref{fig:expsetuplabels0} and Fig.~\ref{fig:expsetuplabels}\subref{fig:expsetuplabels12}.
The feed-forward from Alice's outcome to Bob's measurement choice is crucial in the instrumental scenario, and is experimentally implemented through an optical switch connected to the detector corresponding to Alice's outcome 1. 
The employed optical switch is not ideal, meaning that $5\%$ to $15\%$ of the time (depending on the input rate), it fails to direct the incoming photon to the measurement setting indicated by Alice's feed-forward; in these cases, Bob performs a measurement different from the desired one, so reducing the violation of the CF inequality.

\section{Relaxed tracking and changes in CF inequality bounds}
\label{app:relaxationofAOE}
The assumption in Eq.~\eqref{eq:aoe2} certifies the consistency of Charlie's outcome $c$ with Alice's outcome $a$ when $x = 0$. This is called the tracking assumption and is one of the defining features of the CF correlation set:
\begin{equation}\label{eq:aoeappendix}
    P(a|c,x=0)= \delta_{a,c}.
\end{equation}
We can consider relaxations of the condition Eq.~\eqref{eq:aoeappendix} by allowing non-perfect correlation between $a$ and $c$ bounded up to some value $\epsilon$ thus requiring instead that: 
\begin{align}\label{eq:e-aoe}
\begin{array}{ll}
   P(a|x=0,c) \ge 1 - \epsilon    & \textrm{when} \hspace{0.3cm} a=c  \\
     P(a|x=0,c)  \leq \epsilon   & \textrm{when} \hspace{0.3cm} a\neq c.
\end{array}
\end{align}

Here $\epsilon$ is a measure of the inconsistency of Alice's and Charlie's outcomes: for instance, $\epsilon = 0$ recovers the usual condition of Eq.\eqref{eq:aoeappendix}, while $\epsilon = 1/2$ implies that their outcomes can be completely uncorrelated.  In our analysis, having $\epsilon >0$ generally allows for the possibility of having a nonclassical channel connecting the source with Alice, also when $X=0$. Adopting this assumption leads to a set of $\epsilon$-Relaxed Nonclassical Causal Friendliness ($\epsilon-$RCF) correlations, of which CF correlations are clearly a subset.

It is interesting to see how well a violation of the CF bound compares with this larger set of $\epsilon-$RCF correlations; among other things, it allows us to relax the requirement that Charlie communicates his outcome to Alice with perfect consistency. Knowing how far we are from the $\epsilon-$modified CF inequality also gives us an upper bound on the $\epsilon-$robustness of the experimental results, that is, an estimate of the maximal deviation of the tracking constraint in Eq.~\eqref{eq:aoe2} permissible to warrant a genuine violation of the $\epsilon-$RCF inequality. We will show that when $\epsilon$ is small, the shift in the bound of the CF expression
\begin{align}
    I_{\text{inst}} = \langle AB \rangle_1-\langle AB \rangle_0+ \langle B \rangle_0 - \langle B \rangle_2  \overset{CF}{\le} 2
\end{align}
is small as well, in the sense of the theorem below: 
\begin{theorem}
   \label{Theorem:EpsilonBoundTheorem} For $\epsilon \leq 1/2$, a tight upper bound of $I_{\mathrm{inst}}$ for $\epsilon$-RCF behaviours is 
    \begin{align}
        I_{\mathrm{inst}} \overset{\epsilon-\mathrm{RCF}}{\leq} 2+2\epsilon.
    \end{align}
\end{theorem}

First, let us mention that measures of ``non-absoluteness'' in the context of Local Friendliness constraints over the Bell-scenario-based extensions of the Wigner's friend scenario (sometimes termed canonical Local Friendliness-scenarios~\cite{haddara2025local}) were studied in detail in Ref.~\cite{moreno2022events}. There, direct $\epsilon-$relaxations of the condition in Eq.~\eqref{eq:aoeappendix} and their implications for the set of admissible correlations were also explored. However, the relaxations they considered were of the form (in our language)
\begin{align}
    P(a=c|x=0) \geq 1-\epsilon.
\end{align}
This form of relaxation is generally \emph{not equivalent} to the form we require in Eq.~\eqref{eq:e-aoe}, which explicitly refers to the conditional marginals $P(a|x,c)$ (instead of the joint $p(a,c|x)$), though both recover the usual criterion when $\epsilon=0$. Indeed, the form of relaxation we consider in Eq.~\eqref{eq:e-aoe} is more closely related to allowing some local degree of indeterminism, or more precisely nonclassicality, as was done by Hall~\cite{Hall2010} and, explicitly, the notion of predetermination (more generally, classicality) up to error $\epsilon$ for subsets of settings in the supplemental material (pages 19-20) of Ref.~\cite{ChiribellasystemsCoexisting2024}. Emphatically, however, those results were in the context of Bell inequalities (and different physical motivations), in contrast to the instrumental-based scenario we consider here. 

We shall next demonstrate a proof of Theorem~\ref{Theorem:EpsilonBoundTheorem}. Our derivation is greatly inspired by that of Ref.~\cite{Hall2010} and we adopt similar notation as was used in the appendix of Ref.~\cite{moreno2022events}. First, let us observe that owing to the linearity in probabilities of the expression $I_{\mathrm{inst}}$, and the fact that $\powerset(a,b|x) = \sum_c P(a,b,c|x) = \sum_c P(a,b|x,c)P(c)$ it is possible to equivalently write $I_{\mathrm{inst}} = \sum_{c}P(c) I^{c}_{\mathrm{inst}}$, where $I^{c}_{\mathrm{inst}}$ denotes an expression of similar form evaluated on the conditional distributions $P(a,b|x,c)$. We will derive a bound on $I^c_{\mathrm{inst}}$ that holds for all $c$, hence by averaging over $c$ also for $I_{\mathrm{inst}}$. 

By writing
\begin{align}
\nonumber
    P(+1,+1|x,c) = \gamma^c_{x} \\ 
    P(a=+1|x,c) = \alpha^c_{x}\\ \nonumber
    P(b=+1|x,c) = \beta^c_{x} \nonumber
\end{align}
and using the fact that $\alpha_x^c = \sum_{b}P(+1,b|x,c)$ and so on,  the full probability distributions $P(a,b|x,c)$ can be written as
\begin{equation}
\begin{split}
    P(+1,+1|x,c) &= \gamma^c_{x} \\
    P(+1,-1|x,c) &= \alpha^c_x - \gamma^c_{x}\\
    P(-1,+1|x,c)&= \beta^c_x - \gamma^c_x \\ 
   P(-1,-1|x,c) &=  1-\beta^c_x - \alpha^c_x + \gamma^c_x\;. 
\end{split}
\label{Eq.LastofgammaDefinition}
\end{equation}
Using these expressions, $I^c_{\mathrm{inst}}$ takes the form 
\begin{equation}
    I^c_{\mathrm{inst}} = 4(\gamma^c_1 - \gamma^c_0) + 2(\alpha^c_0 - \alpha_1^c - \beta_1^c - \beta_2^c) + 4\beta^c_0\;.
    \label{Equation:conditionalInstruExpression}
\end{equation}
Now, for all probability distributions, we always have
\begin{equation}
\begin{split}
 \max\{0,\alpha_x^c+\beta_x^c-1\} \leq  \gamma^c_x \leq \min\{\alpha^c_x, \beta_x^c \}\;,
 \end{split} 
 \label{equation:pmarginalssandwichGamma}
\end{equation}
implying the upper bound 
\begin{align}
\nonumber
    I^c_{\mathrm{inst}} \leq &4\min\{\alpha_1^c, \beta_1^c\} - 4\max\{0, \alpha_0^c+\beta_0^c-1\}  \\ &+ 4\beta_0^c +2(\alpha^c_0 - \alpha_1^c - \beta_1^c - \beta_2^c).
\end{align}
Now note that if $\alpha^c_0+\beta_0^c-1 \leq 0$ then $-\max\{0,\alpha_0^c +\beta^c_0-1\}+\beta^c_0 = \beta^c_0 \leq 1-\alpha^c_0$ and if $\alpha^c_0+\beta_0^c-1 > 0$ then $-\max\{0,\alpha_0^c+\beta_0^c-1\} +\beta_0^c = 1-\alpha_0^c$ so that in either case
\begin{align} \nonumber
    I^{c}_{\mathrm{inst}} &\leq 4\min\{\alpha_1^c,\beta_1^c\} + 4-4\alpha_0^c  +2(\alpha^c_0 - \alpha_1^c - \beta_1^c - \beta_2^c)\\
    & = 4 + 4\min\{\alpha_1^c, \beta_1^c \} - 2(\alpha_0^c +\alpha_1^c +\beta_1^c + \beta_2^c).
\end{align}
Recalling that $2\min\{x,y\} = x+y-|x-y|$ the above means that
\begin{align} \nonumber
   I^c_{\mathrm{inst}} &\leq 4+2(\alpha_1^c + \beta_1^c - |\alpha_1^c-\beta_1^c|) - 2(\alpha_0^c + \alpha_1^c +\beta_1^c + \beta_2^c) \\
   &=4 - 2|\alpha_1^c-\beta_1^c| - 2(\alpha_0^c + \beta_2^c) \\ \nonumber
   &\leq 4 - 2\alpha_0^c\;. \label{Equation:UpperboundFromMaximum}
\end{align}
The upper bound $ I^c_{\mathrm{inst}} \leq 4 - 2\alpha_0^c $ is nontrivial, and recovers the bound $I_{\mathrm{inst}}^c \leq2$ immediately if $\alpha_0^c=1$.   Note that by assumption, this is true for CF correlations for the case $c=+1$; however, the derivation of this bound did not require the use of other causal constraints and hence the bound $I^c_{\mathrm{inst}} \leq 2$ is seen to be valid for all distributions with $\alpha_0^c=1$  independently of whether the $P(a,b|x,c)$ satisfy the instrumental inequalities in particular or not.  The bound is also seen to be close to 2 for values of $\alpha_0^c$ close to 1, which makes it suitable to investigate $\epsilon$-RCF. However, for $\alpha_0^c =0$ (and values close to 0), which is implied by the CF assumptions for $c=-1$, it gives no nontrivial constraints (recovering the algebraic upper bound). It furthermore seems that owing to the lack of symmetry in the expression $I^c_{\mathrm{inst}}$, the same derivation with ${\alpha'^c_0} = (1-\alpha^c_0)$, which would imply the other case, does not lead to a similar conclusion. Fortunately, we can cover the case $\alpha^c_0\simeq 0$ by a different approach. 

Recall the instrumental inequalities, which are valid for all correlations compatible with the generalized causal Markov condition:
\begin{align}
     P(a,+1|x,c) + P(a,-1|x',c) \leq 1 \; \; \forall a,  (x,x') \;.
\end{align}
Using Eqs.~\eqref{Eq.LastofgammaDefinition}, these reduce to the following forms of inequality constraints 
\begin{align}
   \label{equation:instrumentalInequalityFamily1}  \gamma^c_x + \alpha^c_{x'} - \gamma^c_{x'} &\leq 1 \\ \nonumber
   \label{equation:instrumentalInequalityFamily2} \beta^c_x - \gamma^c_x  - \beta^c_{x'} - \alpha^c_{x'} + \gamma^c_{x'}  &\leq 0
\end{align}
for all pairs $x,x'$. In particular
\begin{align}
    \gamma^c_1+ \alpha^c_2 -\gamma^c_2 &\leq 1 \\
    \beta^c_0-\gamma^c_0 -\beta^c_1 -\alpha^c_1 +\gamma^c_1 &\leq 0 \\
    \beta^c_0 -\gamma^c_0 -\beta^c_2 -\alpha^c_2 +\gamma^c_2 &\leq 0
\end{align}
and by adding the above together one gets
\begin{align}
   \label{equation:differenceofGammasFromInstrumentalIneq} 2(\gamma^c_1 -\gamma^c_0) \leq 1 + \alpha^c_1 +\beta^c_1 +\beta^c_2 - 2\beta^c_0\; . 
\end{align}
Substituting above to Eq.~\eqref{Equation:conditionalInstruExpression} yields the upper bound
\begin{align}
    I^c_{\mathrm{inst}} \leq  2 + 2\alpha_0^c \;. \label{equation:UpperboundFromInstruineq}
\end{align}
We see that now for $\alpha_0^c=0$, which by CF assumption holds for the case $c=-1$, the upper bound $I^c_{\mathrm{inst}} \leq  2$ is again recovered.  However, the nontrivial bound in Eq.~\eqref{equation:UpperboundFromInstruineq} holds for all (conditional) correlations obeying the instrumental inequalities, not just those which satisfy CF. Together with \eqref{Equation:UpperboundFromMaximum} they give easily obtainable bounds on $\epsilon$-RCF.  Indeed note first that for $\alpha_0^c \in [0,1/2]$ the bound in Eq.~\eqref{equation:UpperboundFromInstruineq} is more constraining than Eq.~\eqref{Equation:UpperboundFromMaximum}, whereas the opposite holds true for $\alpha^c_0\in [1/2,1]$. The bounds coincide and allow for the maximum value of $I^c_{\mathrm{inst}}$ when $\alpha^c_0=1/2$, so that by averaging over $c$ the maximum of $I_{\mathrm{inst}}$ is seen to be at most 3, which is already less than the algebraic upper bound. On the other hand,  by definition of Eq.~\eqref{eq:e-aoe} $\alpha_0^c \in [0,\epsilon]\cup [1-\epsilon, 1]$ for all $c$, corresponding to the cases $P(a=+1|x=0,c=+1) \geq 1-\epsilon$ and $P(a=+1|x=0,c=-1) \leq \epsilon$, where the latter follows from normalization. Consequently for $\alpha^c_0 \in [1-\epsilon, 1]$ with $\epsilon \leq 1/2 $ the bound in Eq.~\eqref{Equation:UpperboundFromMaximum} is more constraining, and gives $I^c_{\mathrm{inst}} \leq 4-2(1-\epsilon) = 2+2\epsilon$ whereas for $\alpha_0^c \in [0,\epsilon]$ with $\epsilon \leq 1/2$ Eq.~\eqref{equation:UpperboundFromInstruineq} is tighter and likewise gives $ I^c_{\mathrm{inst}} \leq 2+2\epsilon$. Thus, by averaging over $c$, the upper bound
\begin{align}
     I_{\mathrm{inst}} \leq 2+2\epsilon
\end{align}
is obtained for all $\epsilon$-RCF correlations with $\epsilon \leq 1/2$. 

It remains to show that this bound is achievable by an $\epsilon$-RCF model. For this purpose, consider a model which has $\alpha_0^{-1}=\epsilon$ and $P(c=-1)=1$ so that $I_{\mathrm{inst}}=I^{c=-1}_{\mathrm{inst}}$ and by comparison of Eqs.~\eqref{Equation:conditionalInstruExpression} and \eqref{equation:differenceofGammasFromInstrumentalIneq} it is seen that the equality $I^{c=-1}_{\mathrm{inst}}=2+2\epsilon $ holds if Eq.~\eqref{equation:differenceofGammasFromInstrumentalIneq} holds with equality. Additionally, we have to check that: $1)$ the value assignments satisfy the instrumental inequalities \eqref{equation:instrumentalInequalityFamily1}--\eqref{equation:instrumentalInequalityFamily2}, $2)$ the assignments are consistent with the relation to the probability distributions $P(a,b|x,c)$  as in Eqs.~\eqref{Eq.LastofgammaDefinition}.  Consider (see acknowledgements) the set: 
\begin{align}
\begin{matrix}
 \label{equation:claudeSolution}  \gamma^{-1}_0 = 0& \alpha_0^{-1} = \epsilon& \beta_0^{-1}=1-\epsilon  \\
   \gamma^{-1}_1= \epsilon & \alpha^{-1}_1=\epsilon & \beta^{-1}_1=1-\epsilon\\
   \gamma^{-1}_2 = 0& \alpha^{-1}_2=1-\epsilon &\beta_2^{-1} =0\;.
   \end{matrix}
\end{align}

We can verify that the assignment in Eq.~\eqref{equation:claudeSolution} is indeed valid by going through all variations of Eqs.~\eqref{equation:instrumentalInequalityFamily1}--\eqref{equation:instrumentalInequalityFamily2} (with $c=-1)$. By writing $f(x,x') = \gamma^{-1}_x + \alpha^{-1}_{x'}-\gamma_{x'}^{-1}$ into a vector $\vec{f}$ and $g(x,x') = \beta_x^{-1} - \gamma_x^{-1} - \beta^{-1}_{x'}-\alpha^{-1}_{x'} +\gamma^{-1}_{x'}$ into a vector $\vec{g}$ with components ordered by $(x,x') = (0,1), (0,2), (1,0), (1,2), (2,0), (2,1)$. After direct computation, one obtains 
\begin{align}
    &\vec{f} = [0, 1-\epsilon, 2\epsilon,1, \epsilon, 0]^T \\ 
    &\vec{g} = [0, 0, -2\epsilon, -\epsilon, -1, -1+\epsilon]^T \;.
\end{align}
Both families of inequalities, which demand $f(x,x')\leq 1$  and $g(x,x') \leq 0$ for all $x,x'$ respectively,  are seen to be satisfied for all $\epsilon \in [0,1/2]$. Furthermore, since  Eq.~\eqref{equation:pmarginalssandwichGamma} holds for all $x$, we confirm that the solution in Eq.~\eqref{equation:claudeSolution} indeed specifies a valid $\epsilon-$parametrized family of behaviours $\powerset^{\epsilon}(a,b|x)$ which satisfy $\epsilon-$RCF constraints and reach the upper bound $I_{\mathrm{inst}} = 2+2\epsilon$.  For completeness, we write the behaviour explicitly. In this model $\powerset^{\epsilon}(a,b|x) = \sum_{c}P^{\epsilon}(a,b|x,c)P(c) = P^{\epsilon}(a,b|x,c=-1)$ and where $P^{\epsilon}(a,b|x,c=-1)$ is obtained using the solution in Eq.~\eqref{equation:claudeSolution} from Eqs.\eqref{Eq.LastofgammaDefinition} as 


\begin{widetext}
\begin{equation}
\renewcommand{\arraystretch}{1.3}
\begin{array}{ll}
P(+1,+1|x=0,c=-1)=0
& P(+1,-1|x=0,c=-1)=\epsilon \\
P(-1,+1|x=0,c=-1)=1-\epsilon
& P(-1,-1|x=0,c=-1)=0 \\
\hline
P(+1,+1|x=1,c=-1)=\epsilon
& P(+1,-1|x=1,c=-1)=0 \\
P(-1,+1|x=1,c=-1)=1-2\epsilon
& P(-1,-1|x=1,c=-1)=\epsilon \\
\hline
P(+1,+1|x=2,c=-1)=0
& P(+1,-1|x=2,c=-1)=1-\epsilon \\
P(-1,+1|x=2,c=-1)=0
& P(-1,-1|x=2,c=-1)=\epsilon
\end{array}
\label{eq:saturating-behaviour}
\end{equation}
\end{widetext}

\newpage

\end{document}